\documentclass[lettersize,journal]{IEEEtran}
\usepackage{amsmath,amsfonts}
\usepackage{algorithmic}
\usepackage{algorithm}
\usepackage{array}
\usepackage[caption=false,font=normalsize,labelfont=sf,textfont=sf]{subfig}
\usepackage{textcomp}
\usepackage{stfloats}
\usepackage{url}
\usepackage{verbatim}
\usepackage{graphicx}
\usepackage{cite}
\usepackage{hyperref}
\usepackage{multirow}
\usepackage{graphicx}
\usepackage{svg}
\usepackage{caption}
\usepackage{color,soul}
\begin{document}

\title{Visualizing Causality in Mixed Reality for Manual Task Learning: A Study}

\let\oldtwocolumn\twocolumn
\renewcommand\twocolumn[1][]{%
    \oldtwocolumn[{#1}{
    \begin{center}
  \includegraphics[width=.8\textwidth]{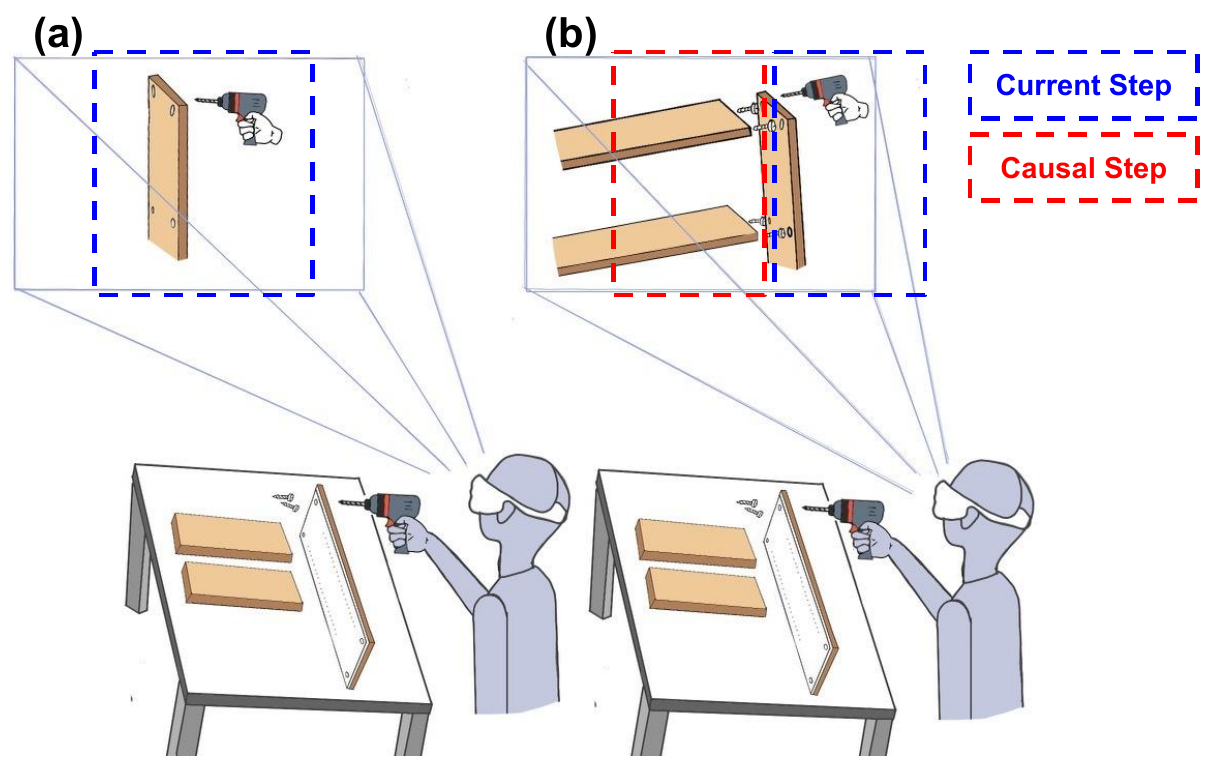}
  \end{center}
   \captionof{figure}{An Illustration of the concept of visualizing causality. Consider an assembly task of a cabinet. Learners can learn the task by perceiving visualization of the steps from the MR device. In (a) learners are only shown the MR instructions of the current step (drill the screws into the plank), while in (b) learners are shown not only the current steps but also the future steps that are causally related (drill the screws into the plank so that it can fit into the other two planks). Our study aims to investigate the effect of showing the causal step in MR manual task learning.}
   \label{fig:teaser}
    }]
}
\author{Rahul Jain*, Jingyu Shi*, Andrew Benton, Moiz Rasheed,
Hyungjun Doh, Subramanian Chidambaram, and Karthik Ramani
\thanks{* These authors contributed equally to this research.}
\thanks{Email: ($jain348|shi537|benton7|mrasheed|hdoh|schidamb|ramani$) $@purdue.edu$}
}

\markboth{Journal of \LaTeX\ Class Files,~Vol.~14, No.~8, August~2021}%
{Shell \MakeLowercase{\textit{et al.}}: A Sample Article Using IEEEtran.cls for IEEE Journals}

\maketitle

\begin{abstract}
Mixed Reality (MR) is gaining prominence in manual task skill learning due to its in-situ, embodied, and immersive experience.
To teach manual tasks, current methodologies break the task into hierarchies (tasks into subtasks) and visualize not only the current subtasks but also the future ones that are causally related.
We investigate the impact of visualizing causality within an MR framework on manual task skill learning.
We conducted a user study with 48 participants, experimenting with how presenting tasks in hierarchical causality levels (no causality, event-level, interaction-level, and gesture-level causality) affects user comprehension and performance in a complex assembly task.
The research finds that displaying all causality levels enhances user understanding and task execution, with a compromise of learning time.
Based on the results, we further provide design recommendations and in-depth discussions for future manual task learning systems.
\end{abstract}

\section{Introduction}
\IEEEPARstart{M}{ixed} Reality (MR) is a technology that combines the physical and digital worlds and has been increasingly utilized for manual task learning across various domains such as assembly~\cite{Whitlock:2020:10.20380/GI2020.43, yamaguchi2020, chidambaram2021processar, chidambaram2022editar, wu2016augmented}, machine tasks~\cite{cao2020exploratory, kong2021tutoriallens, huang2021adaptutar}, and medical training~\cite{eckhoff2018tutar, de2020pre, knudsen2023using}.
Manual task learning involves the acquisition of skills necessary to perform activities that require hand-eye coordination and physical manipulation.
It has thus been significantly enhanced by the introduction of MR applications because of their immersive and realistic scenarios that allow users to practice in a safe and realistic setting with various modalities of instructions.

Current methodologies for manual task learning in MR predominantly focus on guiding the users through the process by visualizing the current steps necessary to perform a particular task~\cite{chidambaram2021processar,kong2021tutoriallens,blattgerste2021train, ajidarma2022skill, stanescu2022model}.
These approaches are designed to support the user's learning process by accurately and quickly guiding users through step-by-step instructions of a task.

 {Derived from these prior works, research pointed out that learners can anticipate their future steps}~\cite{liu2022precueing, ROSENBAUM2007525} {, connected to their current steps.
Similarly, discussion in the psychology community holds that human learns tasks by understanding the causality, i.e., the cause and effect relations among the steps that lie within the tasks}~\cite{gopnik2001causal, kushnir2005young, gopnik2007causal, woodward2005making, woodward2007interventionist}.
 {Enlightened by these findings, researchers have also explored the visualization of causality in various methods such as pre-cueing}~\cite{Wolf2021prevent, liu2022precueing, Liu2021Using, liu2022adaptive}  {or visualizing past or future events}~\cite{fender2022causality, Illing2021Less} { in MR.}

Prior works have focused on diverse applications of visualizing causality in MR and, therefore, enlightened our discussion over the effect of visualization of causality in MR on the human learning process, particularly in our scope, the manual task learning process. 
 {Based on our literature reviews, we categorize the MR visualization of causality in manual tasks into three levels: gesture, interaction, and event.}
 {While prior research has explored the use of MR for manual task learning and the visualization of causality to some extent, no systematic study has examined how different levels of causal visualization: gesture level, interaction level, and event level impacts learning outcomes.
Specifically, there has been no comparative analysis of these methods, leaving a critical gap in understanding how to optimize causal visualization for effective manual task learning in MR environments.}


To this end, we are motivated to contribute to this topic, driven by the belief that a deeper understanding of causality is crucial for more effective manual task learning.
We aim to conduct a study to discover whether and how the visualization of causality in MR affects the understanding of causality in the task and consequently the manual task learning gain.
This study aims to answer the following research questions to reveal future research directions for manual task learning in MR: \textbf{(1) Whether and how does visualizing causality help users learn manual tasks better in MR?} \textbf{(2) How different methods of causal visualization in MR helps users in learning manual tasks effectively?}

To answer these questions, we built a three-level hierarchy for causality representation in manual tasks, deriving from the existing psychology research on causality and human intention~\cite{RIVA201124, pacherie2006towards}, namely \textit{event-level}, \textit{interaction-level}, and \textit{gesture-level}, corresponding to three levels of human-intention and causality.
Based on the derived hierarchy, our embodied demonstrations followed four causal visualization options: \textbf{\textit{no causality}}, {\textbf{\textit{event-level causality}}}, \textbf{\textit{interaction-level causality}}, and \textbf{\textit{gesture-level causality}}, where the users are shown the corresponding demonstrations of manual tasks that are causally connected to their current step. 

To investigate the effects of these four options on the learning gain in manual task learning in MR, we conducted  {a two-phase study} ($N=48$), where we embodied the step-by-step demonstration of the hand-object avatars in a manual assembly task in users' MR view while they were learning the task with the physical components in situ.
The contributions of our papers are as follows:
\begin{itemize}
    \item \textbf{Study Design Rationale and Implementation} of a manual task learning scenario to compare four levels of causality visualization, where \textit{no causality}, \textit{event-level causality}, \textit{interaction-level causality}, and \textit{gesture-level causality} are shown in the demonstration of the manual tasks during users' learning process.
    \item \textbf{Quantitative and Qualitative Results} showing users' objective/subjective responses and learning gains of the manual tasks while following different visualizations of causality.
    \item \textbf{Design Recommendation and In-depth Discussion} summarized from the results and {future opportunity} suggesting promising directions that deepen the research in manual task learning in MR.
\end{itemize}

\section{Related Work}

\subsection{Manual Task Learning in MR}

Mixed reality (MR) integrates physical and digital worlds, creating an environment where physical and digital objects coexist and interact in real-time.
MR is defined as a union of Virtual Reality (VR) and Augmented Reality (AR) by ~\cite{speicher2019mixed}.
MR is gaining interest as a skill acquisition technology due to its uniquely immersive method of delivering educational content ~\cite{dalgarno2010learning}.
When used as a learning platform, MR has shown improved learner achievement, motivation, and attitude towards the learning materials ~\cite{villanueva2021towards, kamarainen2013ecomobile, lu2015integrating}.
Furthermore, embodied interaction, immersion, and situated learning in MR are beneficial in pedagogical contexts~\cite{dalgarno2010learning, martirosov2017virtual, van2013lost, ipsita2022towards} by enhancing learners' spatial abilities ~\cite{lin2015assessing}, enhancing high-level critical thinking ~\cite{saltan2016use}, and reducing the time and cognitive load~\cite{yuan2008augmented}.

These benefits have motivated researchers to investigate and develop MR-based manual-task learning tools and methods.
MR naturally supports spatially-aware instructions for interacting with the physical environment ~\cite{cao2020exploratory}.
Therefore, many task learning methods in MR provide the learner with various designs of spatially-aware, step-by-step instructions by text ~\cite{wojcicki2014supporting,zhu2015context, peniche2012combining}, numerical values~\cite{chardonnet2017augmented, schoop2016drill}, 3D symbolic visual annotations~\cite{funk2016augmented, jo2014unified, kim2019evaluating, westerfield2012intelligent}, and 3D animation of the interactive objects ~\cite{monroy2016mobile, ong2011augmented, webel2013augmented, zhu2014ar}.

Step-by-step instructions by text, annotation, or 3D animation merely explain what to do to the learners, failing to emphasize how to do it.
One method for teaching learners how is by revealing temporal information of position, orientation, and the affordances of the objects along with temporal motion and gesture of the human hands ~\cite{chidambaram2021processar,kong2021tutoriallens,blattgerste2021train, ajidarma2022skill, stanescu2022model}.
Showing concurrent demonstration as guidance in MR to the manual task learners is the most common methodology in existing literature, across diverse domains such as assembly~\cite{Whitlock:2020:10.20380/GI2020.43, yamaguchi2020, chidambaram2021processar, chidambaram2022editar, wu2016augmented}, machine task~\cite{cao2020exploratory, kong2021tutoriallens, huang2021adaptutar}, instrument~\cite{liu2023instrumentar}, physical tasks~\cite{blattgerste2019authorable, thoravi2019loki}, beyond all.
This method focuses on demonstrating the current steps to the users to achieve step-by-step manual task learning.

Based on this prior work, research has noticed the ability of users to anticipate the future steps of the tasks~\cite{liu2022precueing}, and studied the effect of this comprehensive predicting ability~\cite{volmer2018comparison}.
The discussion was open over the effect of demonstration in MR of future and past steps in manual task learning.
We categorize the prior works that involve visualizing cause and effects in MR into three levels: gesture-level, interaction-level, and event-level.
Gesture-level causal visualization reveals the details of hand-object interaction that are required for future steps in a manual task.
Such details involve the location and orientation of an object~\cite{liu2022precueing, Liu2022Rotation}, the hand pose or location~\cite{Wolf2021prevent}, or the hand pose and object pose in a hand-object interaction holistically~\cite{Liu2022testbed}.
Interaction-level causal visualization explains a set of multiple gestures and object movements regarding one object~\cite{liu2022precueing, Liu2022Rotation, Liu2022testbed}.
Interactions depict the dynamic movements of the hands and objects, and, therefore, the temporal similarity among interactions affects the performance of the manual task~\cite{Illing2021Less}.
An event is a set of interactions with different objects.
Many prior works have investigated the effect of replaying causally related past events in MR~\cite{fender2022causality, cho2023realityreplay, bonnail2023memory}, where sequences of past interactions have been captured and visualized to the users to assist in performing the current event.

Despite the coverage of prior research utilizing visualization of causality at different levels in MR, there exists yet no systematic research on the differences of visualizing causality at these levels, nor the discussions over their effects on manual task learning in MR.
To this end, we aim to conduct a study, modeling the causal understanding in manual tasks at different levels, and provide insight into the effect of causal understanding in manual task learning in MR to maximize the learning gain.

\subsection{Causality and Human Intention in Task Learning}

\subsubsection{Definition of Causality}
There's a rich history of philosophical debate on the essence of causality.
Definitions predominantly fall into two types: (1) the abstract of the progress of the world ~\cite{mackie1980cement,whitehead2010process}, and (2) the relationship between human actions and attempts ~\cite{bishop1989natural, mccain1997causal, davis2010causal, clarke2010skilled, aguilar2010causing}.
Similarly, debates are heated over the definition of intention.
Some philosophers ~\cite{honderich2005oxford, setiya2009intention, mele2009intention, borchert2006encyclopedia, craig1998routledge} reduce intention to beliefs and desires to act, while others ~\cite{craig1998routledge, setiya2009intention} see intention as a belief that taking a particular action leads to positively evaluated results ~\cite{craig1998routledge, setiya2009intention, mele1987intentions, roth2000self}.

Since multiple definitions of causality and intention exist, their applicability varies across domains and scenarios.
To ground our work with psychology and philosophy theories that better explain our motivations and framework, we follow the definition of causality and human intention associated with human actions.
By this definition, actions are events that are intentional ~\cite{craig1998routledge, stuchlik2013volitionalism}, i.e., they are caused by human intention.
The events take place necessarily in planned ways in a task~\cite{mclaughlin2009oxford, audi1999cambridge, wilson2012action}. The human's ability to understand and replicate the planned occurrence is the key to learning a task, and we define this planned causality as \textbf{intention-driven causality}.

Although the definition above can be debated, such as the absence of unintentional actions ~\cite{vernon1979unintended, boudon2016unintended} and awareness of unintended side-effects (i.e., oblique intentions) ~\cite{guglielmo2010can}, we argue that in the task learning process, learners will only be taught intended tasks.
Thus the definitions provided serve as a sufficiently working model.

\subsubsection{Causality in Task Learning}
Research in cognitive and developmental psychology indicates the significance of understanding causality and intention in task learning. 
Humans have been found to parse and understand actions from intentions instead of motion from infancy ~\cite{meltzoff2001like, meltzoff1995understanding, george2008caregiving} and early childhood~\cite{astington2001paradox}.
In addition, Heyes et al. ~\cite{heyes2014cultural} state that humans can interpret the intentions of others and generalize this knowledge to predict their actions in novel contexts.

Causality can be inferred via probabilistic learning ~\cite{gopnik2001causal, kushnir2005young} or through one's trials of the task~\cite{gopnik2007causal, woodward2005making, woodward2007interventionist}. However, learning causality by observation has shown to be more effective~\cite{meltzoff2012learning, bonawitz2010just}.
Learners comprehend the cause and effect of each action by observing the demonstrations of instructors/teachers ~\cite{bandura2008observational, douglas2006observational, teo2021learning}.
Such observational causal learning is guided by intention.
When learners apply the task they have learned, with causality in mind, they bind, extend, and generalize the causes and effects of their actions ~\cite{buehner2009causal}.

\section{Design Consideration}

\subsection{Hierarchy of Causality in Task}
Within the domains of Cognitive Science, Psychology, and Neuroscience, scholars have put forth a postulation that humans possess a tendency to segment complex tasks into distinct groups~\cite{barker1951one, michotte2017perception}. Additionally, people understand those ongoing tasks in partonomic hierarchies~\cite{kurby2008segmentation}. 
Similarly, we break the hierarchical task structure into numerous discrete events, which can be further parsed into interactions and subsequently into gestures. 
These entities — events, interactions, and gestures —are henceforth denoted as elements within this hierarchical framework.  
The occurrence of these elements is decided by the intentions at different levels which are P-Intentions (Events), D-Intentions(Interactions), and M-Intentions(Gestures)~\cite{RIVA201124}. 

\begin{figure*}
    \centering
	   \includegraphics[width=.8\textwidth]{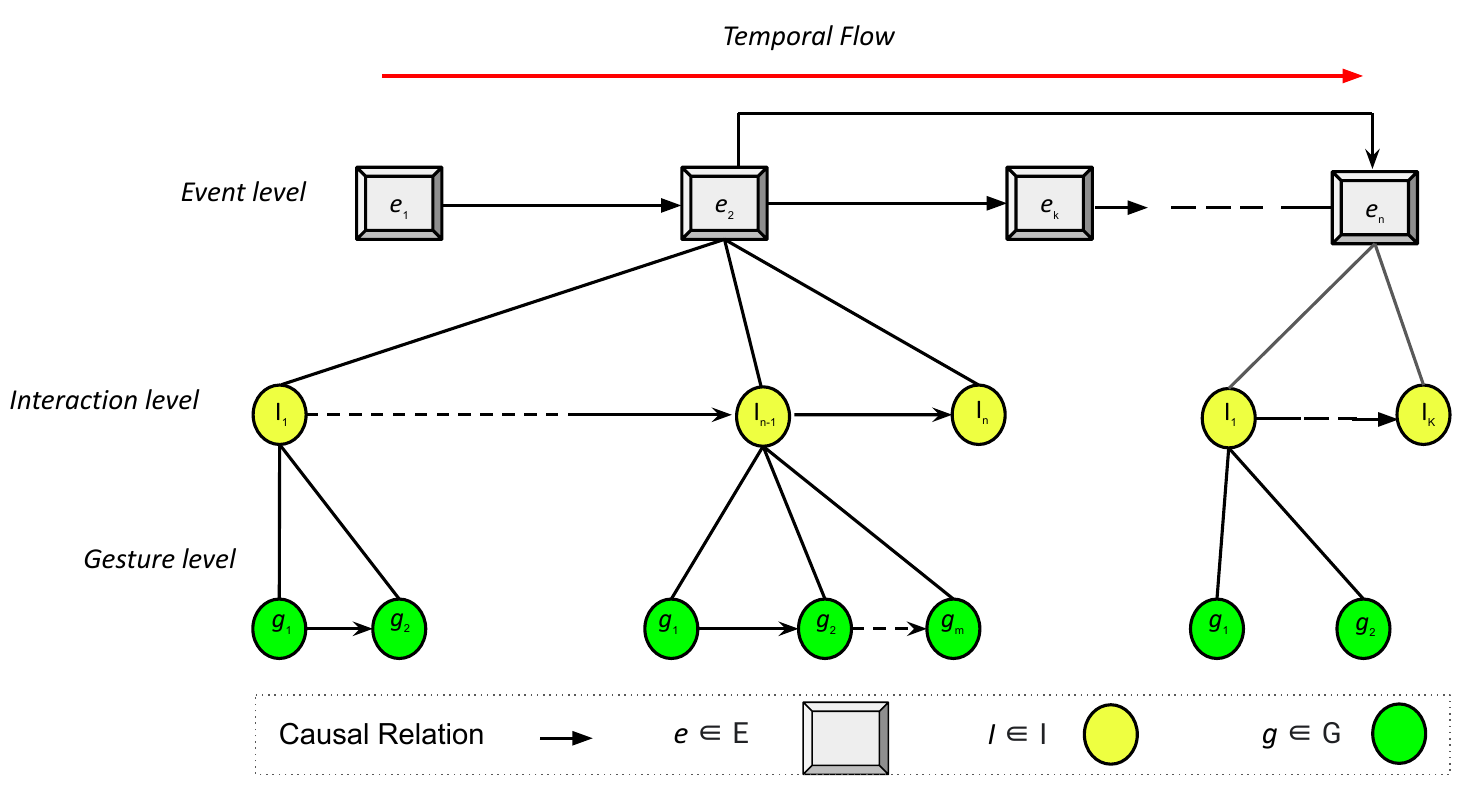}
    \caption{A three-level hierarchy for causality in a manual task. The three levels are event, interaction, and gesture, respectively. Horizontally, each node is connected by causal relations, explaining the cause and effect among them. E.g., the occurrence of event 1 leads to the occurrence of event 2. A node from a higher level consists of one or multiple children at the lower level. We aim to differentiate different levels of causality and study their effects in manual task learning.}
    \label{fig:taxanomy}
\end{figure*}

Moreover, intentions not only decide the occurrences but also decide the order and pattern of the elements in the inter-layer(i.e., they are only connected in the same layer). These inter-layer connections between elements are related by cause and effect(causality)~\cite{michotte2017perception}. This implies that the intentions themselves drive the causality observed at each level of the hierarchy, and this \textbf{intention-driven causality}~\autoref{fig:taxanomy}. includes 1) Event level causality 2) Interaction level causality and 3) gesture level causality. 

\textbf{Event Level Causality:}
This refers to causal links between events. Events are connected in a temporal sequence, often involving cause-and-effect relationships. The causality between events is driven by intention, particularly the intention to complete a task.

\textbf{Interaction Level Causality:}
This type of causality focuses on the relationships between interactions within an event. Interactions could be more specific actions, behaviors, or steps that contribute to the completion of an event. The causal relations between interactions are established due to the intention set at the event level. The overall aim here is to accomplish the event successfully.

\textbf{Gesture Level Causality:}
Gesture-level causality deals with the temporal links between poses (referring to physical postures, configurations, or conditions) within an event. These causal relations are influenced by the intentions set at the interaction level. In this context, poses represent more granular components of interactions, and their causality is aimed at fulfilling the requirements of the interactions.

 {For example, consider the task of assembling furniture. At the \textbf{event level}, the task could involve "assembling the frame," "attaching the shelves," and "tightening the screws." These events must occur in a specific sequence, the frame must be assembled before attaching the shelves.
The causality here is driven by the assembler’s P-Intention to complete the furniture assembly.

At the \textbf{interaction level}, within the event of "assembling the frame," interactions could include "positioning the side panels," "aligning the bolts," and "inserting the bolts into the frame."
These interactions are causally linked, positioning the side panels must occur before aligning and inserting the bolts.
The assembler’s D-Intention to successfully complete the frame event governs this level of causality.

At the \textbf{gesture level}, during the interaction of "inserting the bolts," gestures such as "grasping the bolt," "aligning the bolt with the hole," and "turning it with a wrench" come into play.
These gestures are causally connected, where misalignment in one gesture can disrupt the interaction.
These gestures are guided by the M-Intention to complete the interaction correctly.}

\subsection{Visualization}

Effective task learning often involves the emulation of an expert practitioner, where learners observe and replicate demonstrations provided by a proficient master\cite{cao2020exploratory}. These demonstrations are typically recorded and delivered to the learners. Prior research has emphasized the importance of visualizing task hierarchies in a Mixed Reality environment\cite{blattgerste2019authorable, thoravi2019loki, huang2021adaptutar}.  
We follow similar approach for our study in which we present the visualization of the task by showing the hierarchies using a a graph - Events, Interactions, and Gestures in a Mixed reality environment. The graph structure involves three nodes for events, interactions, and gestures as shown in ~\autoref{fig:visualization}  

\subsubsection{Causality Visualization}
\begin{figure}
    \centering
	   \includegraphics[width=.4\textwidth]{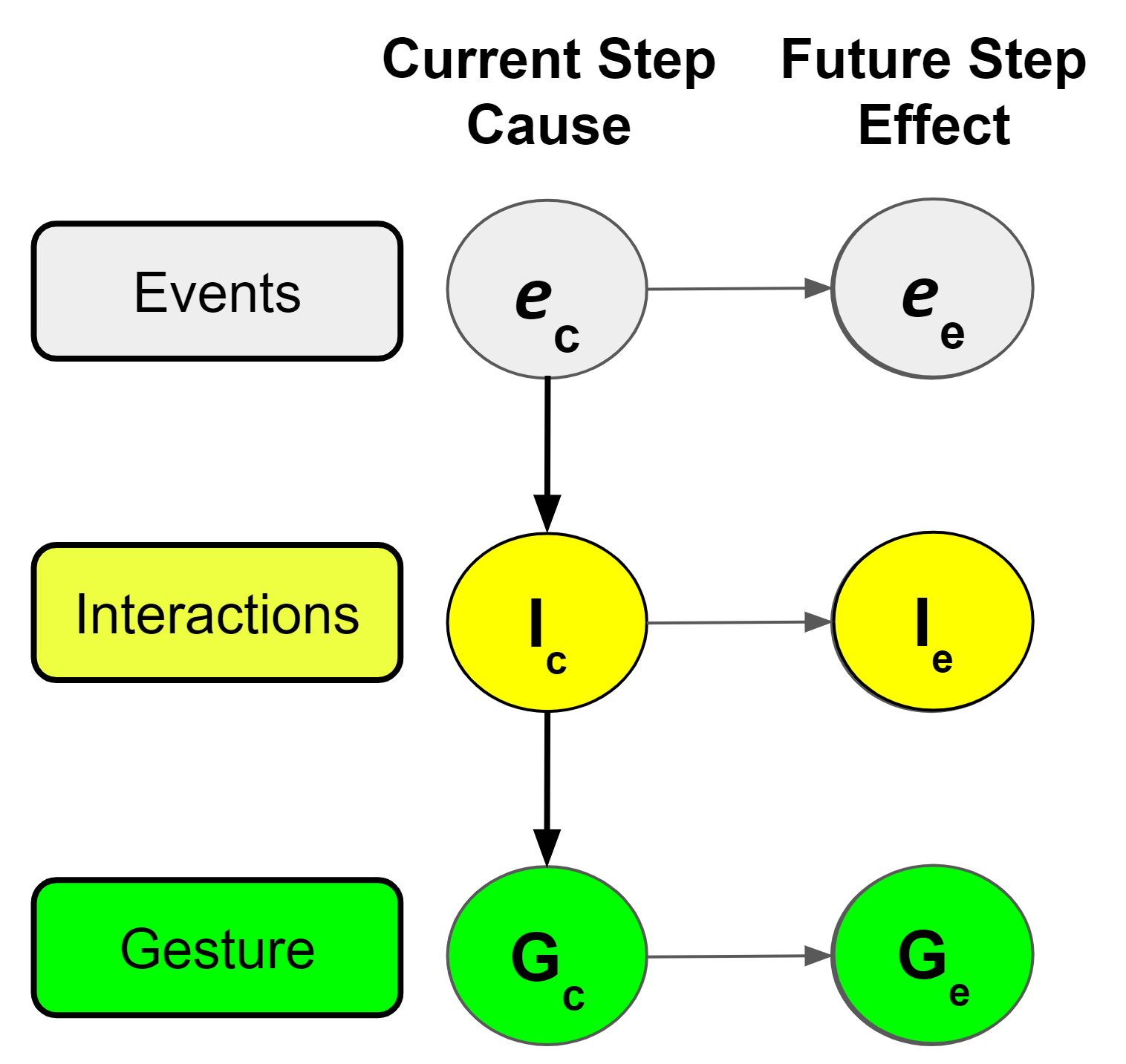}
    \caption{Our methodology to visualize the causality in the manual task at three levels. In the Cause column, we will show the demonstration of (event, interaction, or gesture of) the current step, which is the cause. In the Effect column, we will show the corresponding level of demonstration of the future step that is causally related to the current step.}
    \label{fig:visualization}
\end{figure}

There are other tutoring systems that provide information on the current task as well as the future task in AR~\cite{liu2022precueing}. In a similar way instead of showing the next task, we show the causal task (effect of the current tasK) to the learner. Other prior works have used textual and visuals~\cite{fender2022causality} narrative with graphs to help users understand temporality in the task. We prefer textual and visual information to show causal graphs because visuals stick to memory more effectively. For our study, We show current event, interactions, and gesture and their effect (the future) together in the mixed reality in the form of a graph as shown in Figure~\ref{fig:visualization}.    

\subsubsection{Demonstration of Content}
When a user is trying to learn, the most effective way is to observe and follow the demonstration of an experienced master. The expert records the content and delivers it to the user. Similar to prior work\cite{chidambaram2021processar} which allows both video and demonstration. We further provide the users with the functions to see the 3D hand-object-interaction demonstrations of the manual tasks in MR. This allows users to see instructions as well as demonstrations in MR.

\section{Test Bed}
We implemented an MR scene to evaluate the effect of causal visualization in learning tasks.
The user interface and demonstrations of the content were developed in Unity software.

\begin{figure*}[htb]
  \centering
  \begin{minipage}{.6\textwidth}
    \centering
    \includegraphics[width=\linewidth]{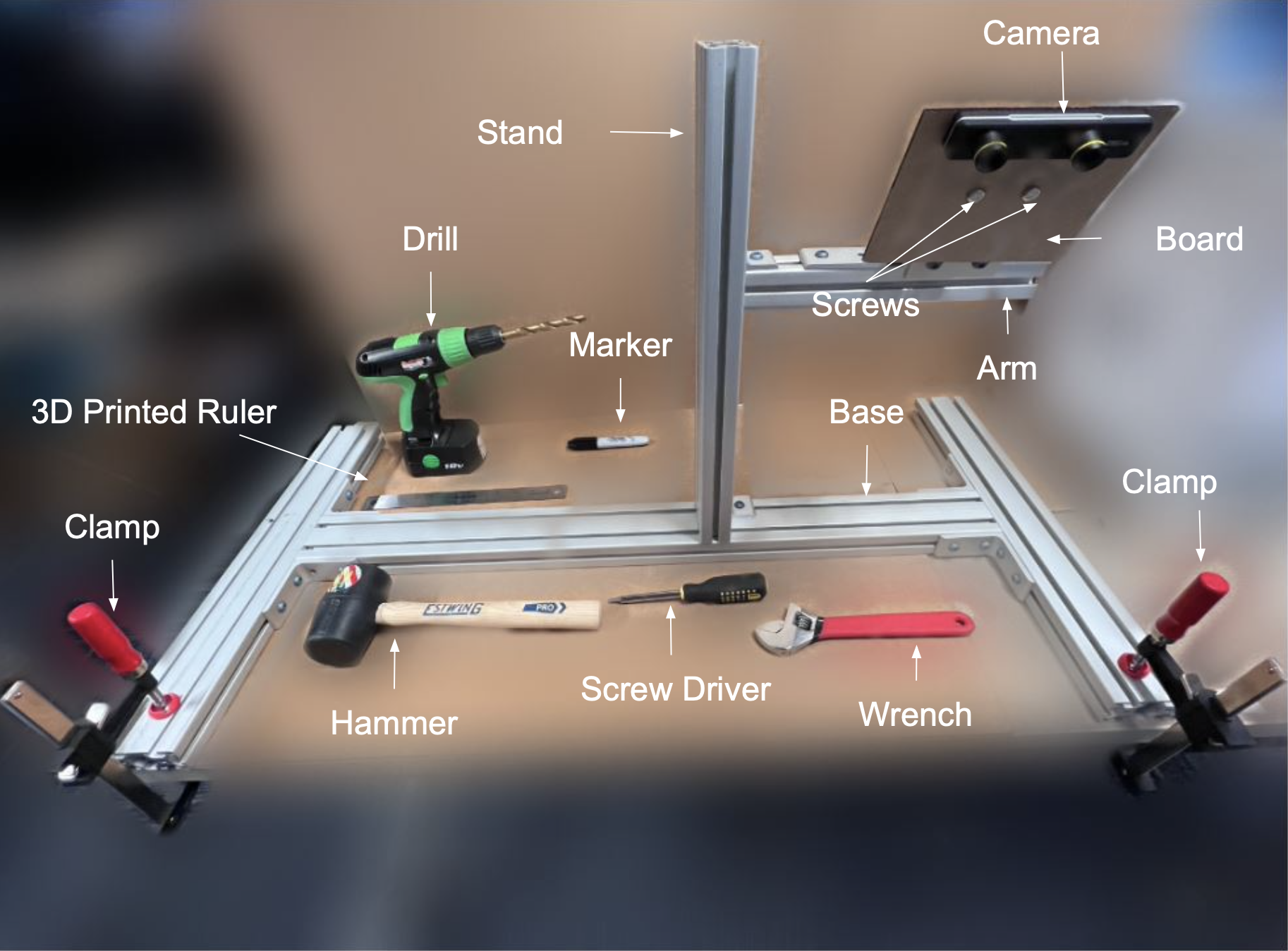}
    \caption{The camera setup to be assembled and the tools to be used in our test bed.}
    \label{fig:camera_setup}
  \end{minipage}%
  \hfill
  \begin{minipage}{.32\textwidth}
    \centering
    \includegraphics[width=\linewidth]{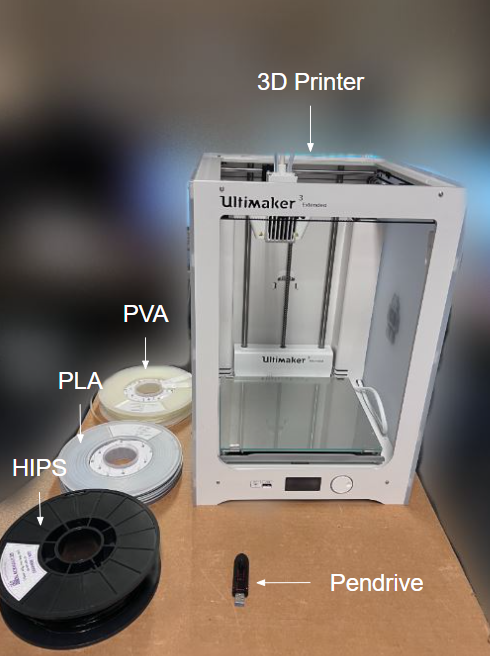}
    \caption{The printer setup in our test bed.}
    \label{fig:printer_setup}
  \end{minipage}
\end{figure*}

\subsection{Hardware Implmentation}
We developed an MR system by attaching the ZED Mini stereo camera to the Oculus Quest 2.
The system uses the ZED Mini to capture both the physical scene and depth information of the environment while the Quest 2 gathers hand-pose information and displays the visualization and demonstration in the MR scene.
The creation of virtual content and hardware integration was done using the Unity [41] game engine.

\subsection{Test scene} 
In our study, we used the task of assembling a camera setup \autoref{fig:camera_setup} as a test scene.
The task was chosen because of its procedural nature and complexity (i.e. it is hard to perform the task without looking at instructions).
The task supports all the elements in the hierarchy which are events, interactions, and gestures.
The task consists of five events: (1) Move the base onto a table, fix the base with two clamps, and check that it’s set by striking it with the hammer.
(2) Attach materials to a 3D printer, insert a flash drive, and press the start button to print a ruler.
(3) Attach a vertical beam to a designated location on the base and fasten it using a screw and screwdriver.
(4) Attach an arm to a designated location on the vertical beam and fasten it using a screw and screwdriver.
(5) Attach a camera mount to the arm and fasten it with a screwdriver. Then, drill two holes in a wooden board and use two bolts to secure the board to the mount.
The detailed procedure of the task is present in \autoref{tab:my_label}. 

\begin{table*}[htp]
    \centering
    \caption{Events and interactions of the manual assembly tasks with our test bed.}
    \includegraphics[width=\linewidth]{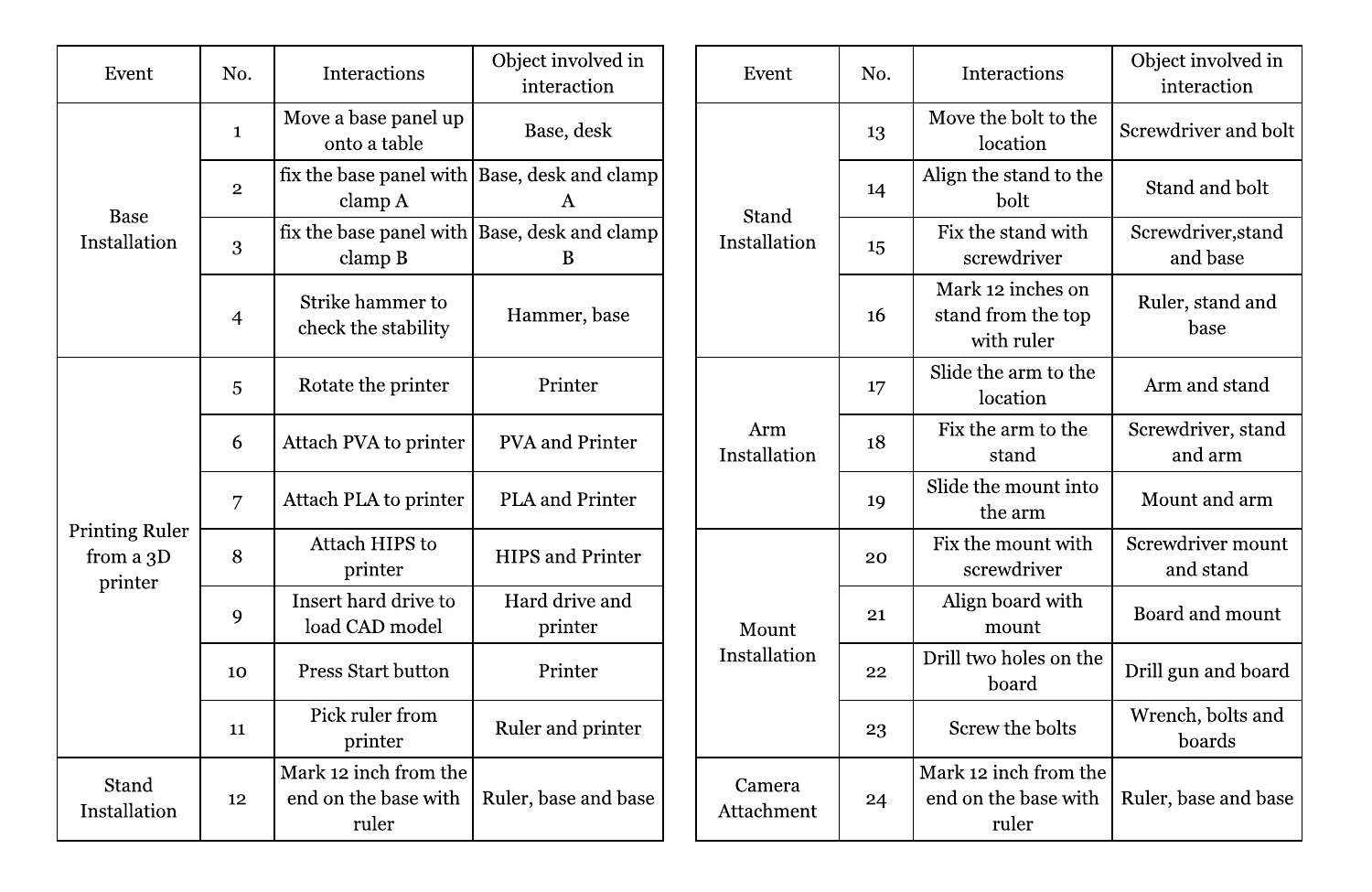}
    \label{tab:my_label}
\end{table*}

\subsection{Authoring Content}
We developed a system that converts live demonstrations to 3D content for generating tutorials for the learner. 
To create 3D tutorials of the camera setup, CAD models of the objects were created through scanning by an Intel RealSense 435i camera.
The camera was moved 360 degrees around the object to capture RGB-D frames which were used by BADSLAM~\cite{schops2019badslam}, an RGB-D SLAM method to generate a 3D mesh.
The 3D models for all objects involved in the camera setup were imported to Unity.
Then one of the authors wears an Oculus 2 with a zed camera enabling passthrough to create the content.
In the beginning, the scanned mesh models are aligned manually with the respective physical object.
This alignment provides the initial six degrees of freedom (6DoF) pose of the objects. After that author performs the entire procedure.
To create 3D content, hand tracking, and object 6 Dof tracking are needed.
During the demonstration, Hand-tracking data and RGB-D frames were collected along with RGB frames from the Zed Camera and Oculus 2.
The collected RGB and RGB-D frames were used to track object 6DoF using Megapose~\cite{labbe2022megapose}.
Both 3D hand tracking data and object 6Dof data were used for creating 3D content for the entire camera setup tutorial. 

To find the elements of the camera assembly task, we manually segment the entire 3D content into five events.
To segment interactions in each event, we temporally segment the events into interactions.
Each interaction consists of four stages: (1) hand approaching an object, (2) grasping the object, (3) manipulating the object, and (4) releasing the object.
For finding pose, we only consider gesture used in stage 2 when the hand is grasping the object.
Causal links between the elements are also manually established using the same method used by Fender and Holz, (2022)~\cite{fender2022causality}.
These causal links and the order of operations generate our learning graph for the task of authoring content.

\section{User Study}
\label{userstudy}
 {We conducted a user study with our test bed to evaluate two evaluate our research questions. 
In the first, we studied the impact of the existence of causality visualization on the learning process of a manual task.
In the second, we studied the impact of different levels of causality visualization on the learning process.}

\subsection{Hypotheses}
 {Based on our research questions and literature review, we formulated following hypothesis based on time and errors which are  well-established factors in assessing performance in manual task learning.} \\
For  {RQ} 1:
\begin{itemize}
    \item \textbf{H1.} \textit{Participants will spend less time learning the tasks if shown the demonstrations of steps of the task that are causally related to their current step than if shown only the demonstration of their current task.}
    \item \textbf{H2.} \textit{Participants will perform faster in the task after they learn the task by understanding the causality in the task.}
    \item \textbf{H3.} \textit{Participants will make fewer mistakes in the task after they learn the task by understanding the causality in the task.}
\end{itemize}
For  {RQ} 2:
\begin{itemize}
    \item \textbf{H4.} \textit{Among three levels of causality, participants will spend the least time learning if shown the gesture-level causality, and the most time if shown the event-level causality.}
    \item \textbf{H5.} \textit{Among three levels of causality, participants will perform fastest after they learn the gesture-level causality of the task and slowest after they learn the event-level causality.}
    \item \textbf{H6.} \textit{Among three levels of causality, participants will make the least mistakes in the task after they learn the gesture-level causality of the task and the most mistakes after they learn the event-level causality.}
\end{itemize}

\subsection{Study Design}
\label{Study Design}
\subsubsection{Participants}
\begin{figure*}[ht]
  \centering
  \includegraphics[width=1\linewidth]{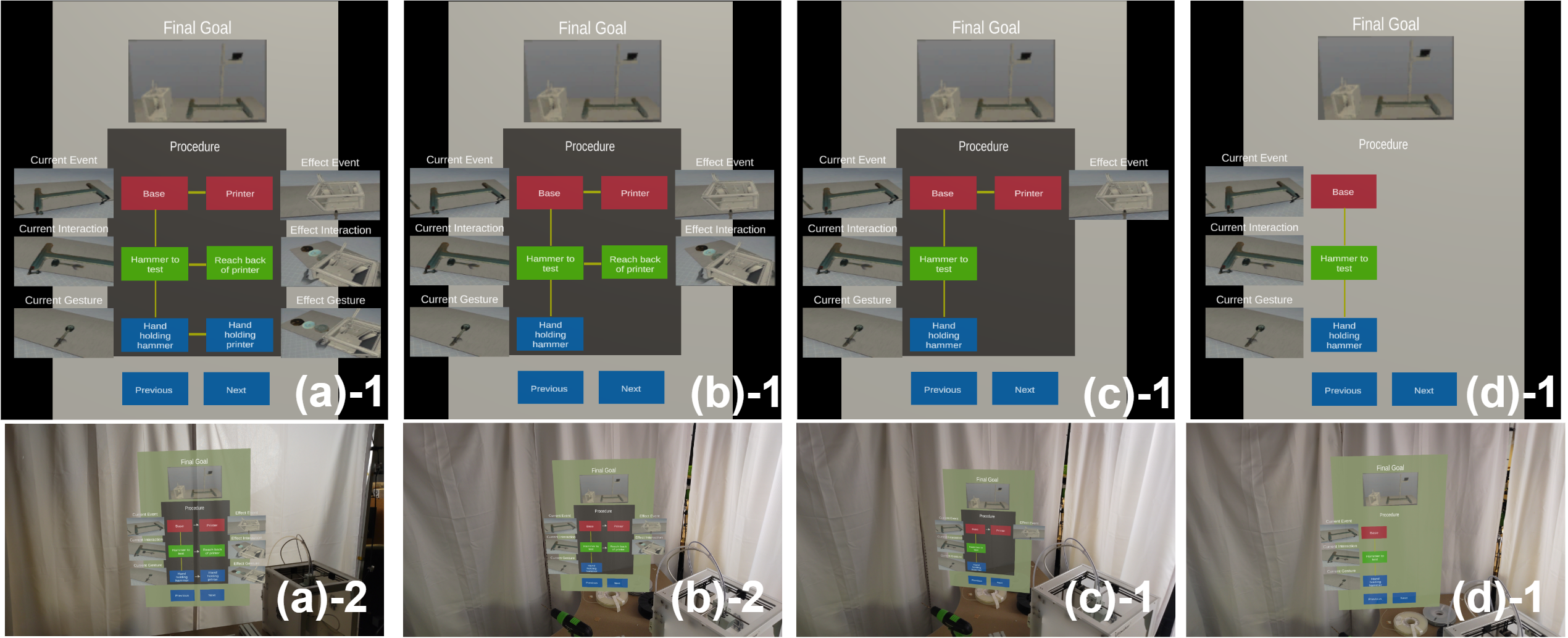}
  \caption{UIs that were shown to 4 different groups and AR animation. a) shows the event, interaction, and affordance/gesture, b) shows the event and interaction level relation, c) shows the event level causal relation and finally d) shows the graph and no causal relation.}
  \label{fig:interface}
\end{figure*}

We recruited 48 users (25 self-identified as males and 17 self-identified as females, aged 18 to 55) for our user study.
Among the user group, 14 have used either VR or AR devices before, 12 have used both before, and 20 have never used either one.
Thirty-six users have a background in engineering, eight in science, one in education, and three in others.
Each participant was compensated with a \$15 gift card.
All participants completed the entire study.
We divided our users into four groups, each provided with different levels of information from the learning graph. 
\begin{enumerate}
    \item The Control Group {\textbf{GC}} were shown 3D demonstrations of the hand gestures, object affordances, interactions, and events of the current step~\autoref{fig:interface} (d).
    \item The First Group {\textbf{G1}} were shown 3D demonstrations of the hand gestures, object affordances, interactions, and events of the current step. Meanwhile, they are shown demonstrations of the steps that are causally related at the event level~\autoref{fig:interface} (c).
    \item The Second Group {\textbf{G2}} were shown 3D demonstrations of the hand gestures, object affordances, interactions, and events of the current step. Meanwhile, they are shown demonstrations of the steps that are causally related at the interaction and event levels~\autoref{fig:interface} (b).
    \item The First Group {\textbf{G3}} was shown 3D demonstrations of the hand gestures, object affordances, interactions, and events of the current step. Meanwhile, they are also shown demonstrations of the steps that are causally related at all three levels~\autoref{fig:interface} (a).
\end{enumerate}

We evaluated the learning performance by modeling the causality for Groups (\textbf{\textbf{G1, G2}, and \textbf{G3}}), while the Control group (\textbf{\textbf{GC}}) was provided no information on causality.
This approach enabled comparison of the learning performance between participants exposed to different levels of our taxonomy for modeling causality.
 {Testing causality levels hierarchically mirrors how learners typically develop their understanding level by level.
This approach allows us to evaluate how incremental layers of causality improve task learning, rather than isolating them as independent conditions.
Therefore, we chose to test causality levels hierarchically instead of independently.}
In our study, we would like to clarify that the understanding of causality has a cliff-like learning effect, i.e., "know if a thing will happen or not," is an approximate binary variable.
We design our user study with a between-subject setup.

\begin{figure}[htp]
    \centering
    \includegraphics[width=.8\linewidth]{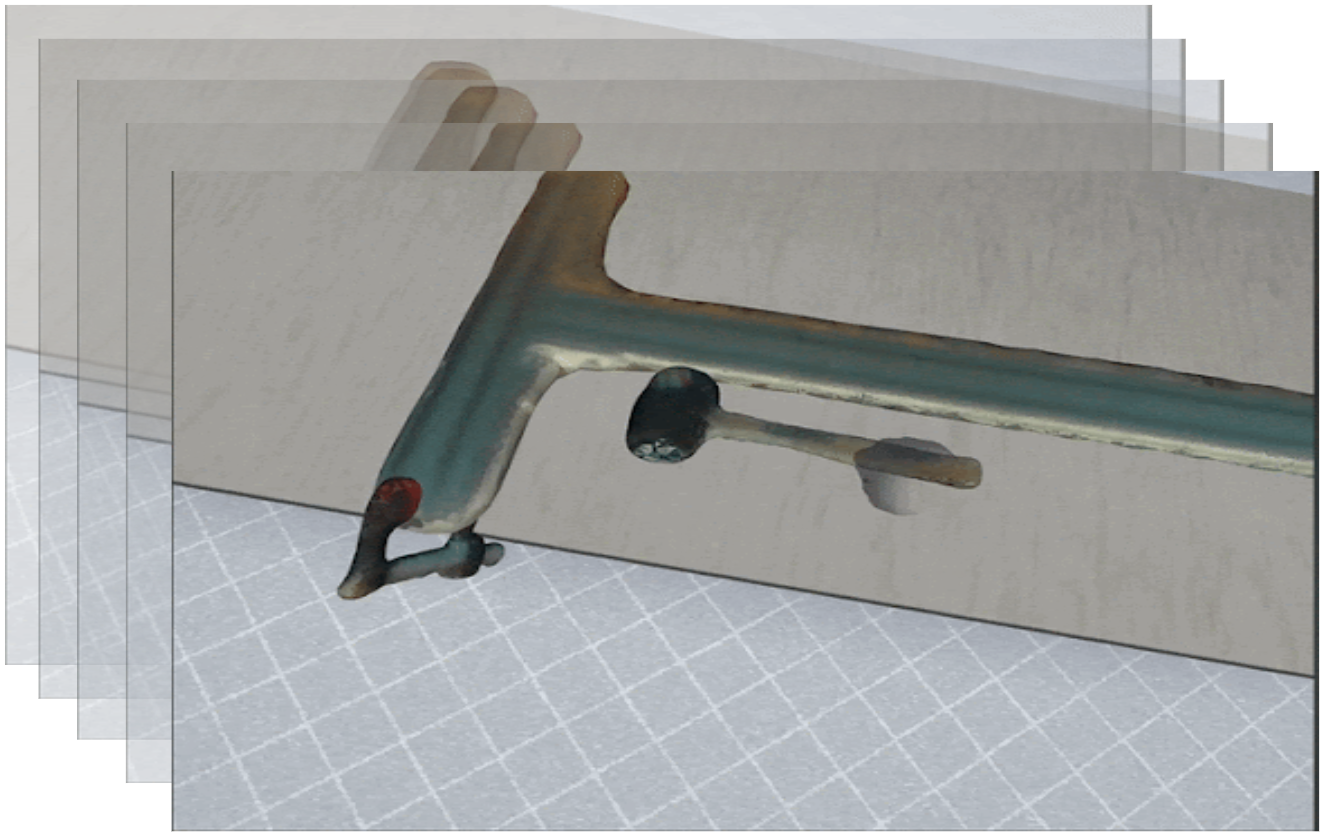}
    \caption{The VR content shown to the user in the visualization graph.}
    \label{fig:content}
\end{figure}

\begin{figure}[htp]
    \centering
    \includegraphics[width=.8\linewidth]{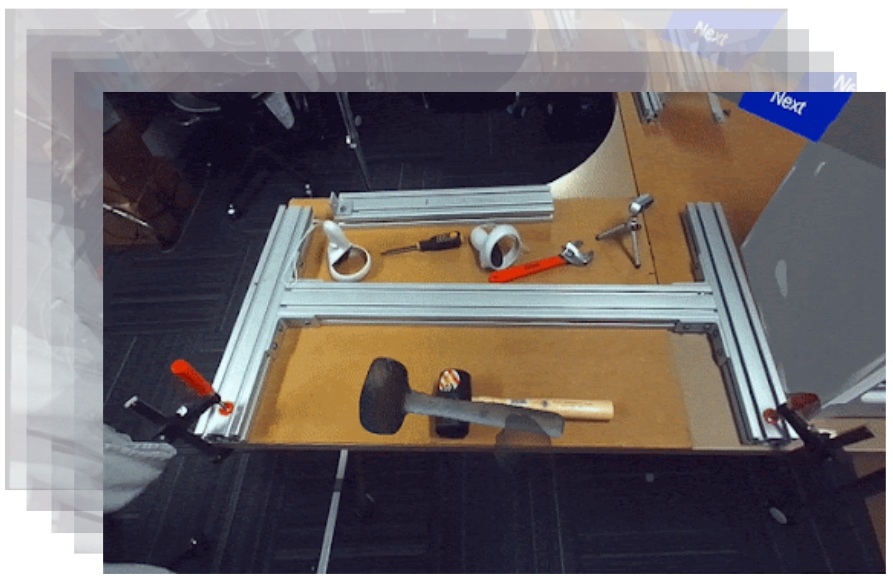}
    \caption{AR demonstration of the current interaction.} 
    \label{fig:demo_AR}
\end{figure}
\subsubsection{Procedure}
Each participant was warmly welcomed by the researcher and provided with a concise introduction to both the study and the task.
 {The study was conducted under the approval of the Institutional Review Board.}
Upon signing the consent form, participants received a brief overview of the task's details.
Then, they were asked to wear headsets with pass-through capabilities to get familiar with our system in the MR environment.
The study includes two phases: the learning phase and the testing phase.
In the learning phase, each user was provided with sufficient time (1 hour) to learn the content using our system.
Every participant finished learning within one hour. Then, participants performed the entire assembly task without any help in the testing phase. This phase has a 15-minute time limit to complete the task.
 {The pre-authored content is visualized in the UI panel in the form of VR scene snapshots as exemplified in} ~\autoref{fig:content}.
 {Participants can also see AR demonstrations of the current interaction as exemplified in}~\autoref{fig:demo_AR}.
After finishing the study, participants provided feedback about the study through a semi-structured interview and questionnaires.

\subsubsection{Data Collection}

 {Throughout the study, we recorded participants' performances from the third-person camera as well as a screen capture of the AR view.} 
We then manually computed the length of time of each  {phase}, event, and interaction alongside task/sub-task accuracy in post-processing.
We visually compared the procedural consistency between participants' demonstrations of tutorials at all three levels.
We analyzed the number of times each participant from \textbf{Gc} and \textbf{G3} revisited the instruction, i.e., they referred to the instruction again for more information.
We did not include participants from \textbf{G1} and \textbf{G2} since the coverage in the length of the instructions is different. Therefore, it is of no need or rigor to compare with them.
After the study, participants provided feedback on our system through the NASA TLX~\cite{nasa1986task} form and a semi-structured interview, 
Participants also completed a 5-point Likert scale questionnaire.
{The questions from the survey were designed to determine the participant's memory of the task, self-rating of their memory from learning, motivation to learn, and comprehension of the causality and intention within the tasks}~\cite{alarid1976investigation, lindwall2012instruction}, {we designed the questionnaires following prior works on learning manual tasks in MR}~\cite{huang2021adaptutar, ipsita2022towards, chidambaram2021processar}.

\section{Results and Data Analysis}
This section evaluates our workflow by analyzing participants' performance, subjective feedback, and other interesting findings.

The independent variable manipulated between our four groups is the level of causality being visualized in the learning system.
As described in ~\autoref{Study Design}, {Gc} saw no intent and causality information while {G1},{G2}, and {G3} were exposed to various levels of our taxonomy.
Note that the level of causality (being visualized) is defined by the cause-and-effect pair of the level, hence both should be visualized accordingly.
The dependent variables for each of the hypotheses in our experiments are time spent in learning (\textbf{H1, H4}), time spent completing the task after learning (\textbf{H2, H5}), and the mistakes made while completing the task (\textbf{H3, H6}).

Thus, we make four comparisons ({Gc} vs {G3}, {G1} vs {G2}, {G1} vs {G3}, and {G2} vs {G3}). 
It is shown by the Shapiro-Wilk normality test ($p<0.005$) that the data from \textbf{neither} our quantitative nor qualitative evaluation follows a normal distribution.
Therefore, we conducted a Friedman test for {Gc} vs {G1} vs {G2} vs {G3} and a Wilcoxon signed-rank test for all four comparisons individually. 
To prove the hypothesis of  {RQ} 1, we compare {Gc} vs {G3} and for  {RQ} 2 the other three comparisons. 
\subsection{Objective Performance}

\subsubsection{Learning Performance}

\begin{figure}[htp]
    \centering
	   \includegraphics[width=\linewidth]{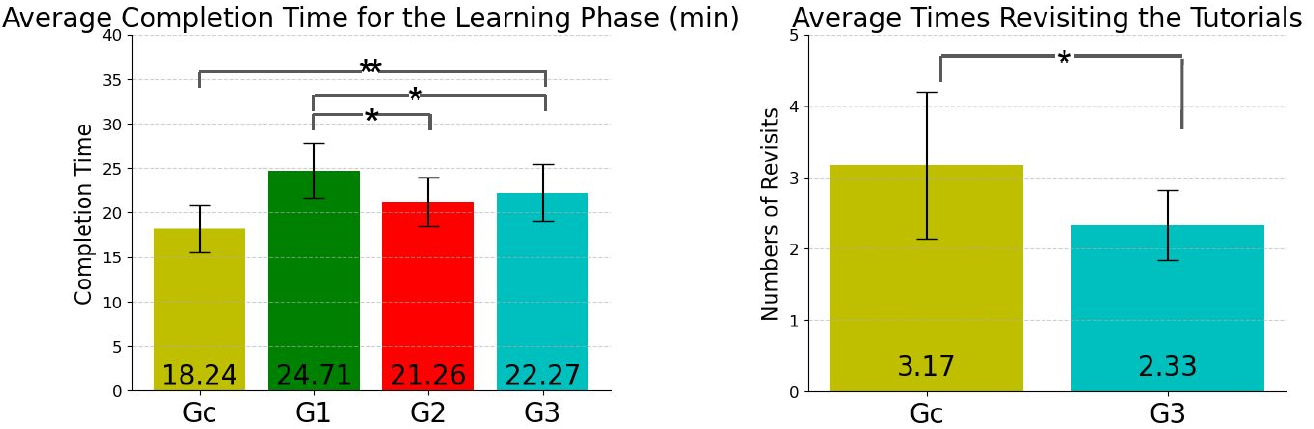}
    \caption{Objective Results from Learning phase. The comparison of \textbf{average learning time} is on the left-hand side. The comparison of the \textbf{average number of revisits in the instructions} is on the right-hand side. ($**=p<.005,*=p<.05$, error bars represent standard deviations)}
    \label{fig:learning_results}
\end{figure}

We first demonstrate the overall learning performance by comparing the total time spent in the learning phase, including viewing the content and practicing.
The results are shown in ~\autoref{fig:learning_results}.

The Friedman test shows significant differences among the groups in learning time($\chi^2(3)=18.725, p=0.003$).   
The average learning time shows that the participants spend the longest amount of time (all in minutes) when provided event-level causal demonstration in {G1}($M=24.71, SD=3.08$).
To evaluate \textbf{H1}, we compare the {Gc} and {G3}.The average learning time of participants from {G3} ($M=22.27, SD=3.23$) spend more time than those from {Gc} ($M=18.24, SD=2.60, p=0.001, Z=-2.981$). This shows that {G3} which includes causal information increases the learning (training) time as compared to {Gc} which shows only the current task. Hence \textbf{H1} is failed.   
For \textbf{H4}, we compare among {G1}, {G2} and {G3}.
 {The time spent in learning of {G2} ($M=21.26, SD=2.71$) is not significantly different from that of {G3} ($p=0.140, Z=-1.098$).}
Meanwhile, participants from {G2} spend significantly less time in learning than those from {G1} ($p=0.008, Z=-2.401$).
Participants from {G1} also spent significantly more time learning than those from {G3} ($p=0.050, Z=-1.647$). We conclude that \textbf{H4} is partly proved as the most time was taken by {G1} and least by {G2}

\subsubsection{Testing Performance}

\begin{figure}[htp]
    \centering
    \includegraphics[width=.45\textwidth]{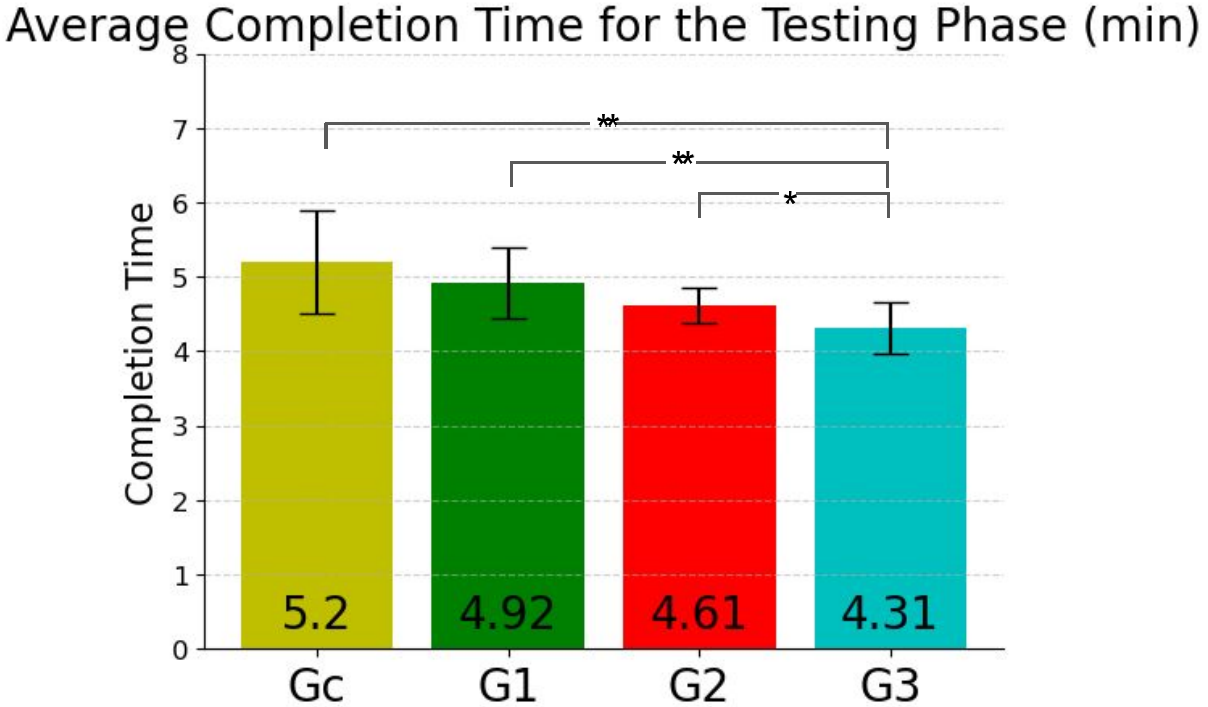}
    \caption{\textbf{Average time (in minutes) spent in the testing phase} by groups. ($**=p<.005,*=p<.05$, error bars represent standard deviations)}
    \label{fig:testing_time}
\end{figure}

\begin{figure}[htp]
    \centering
    \includegraphics[width=.45\textwidth]{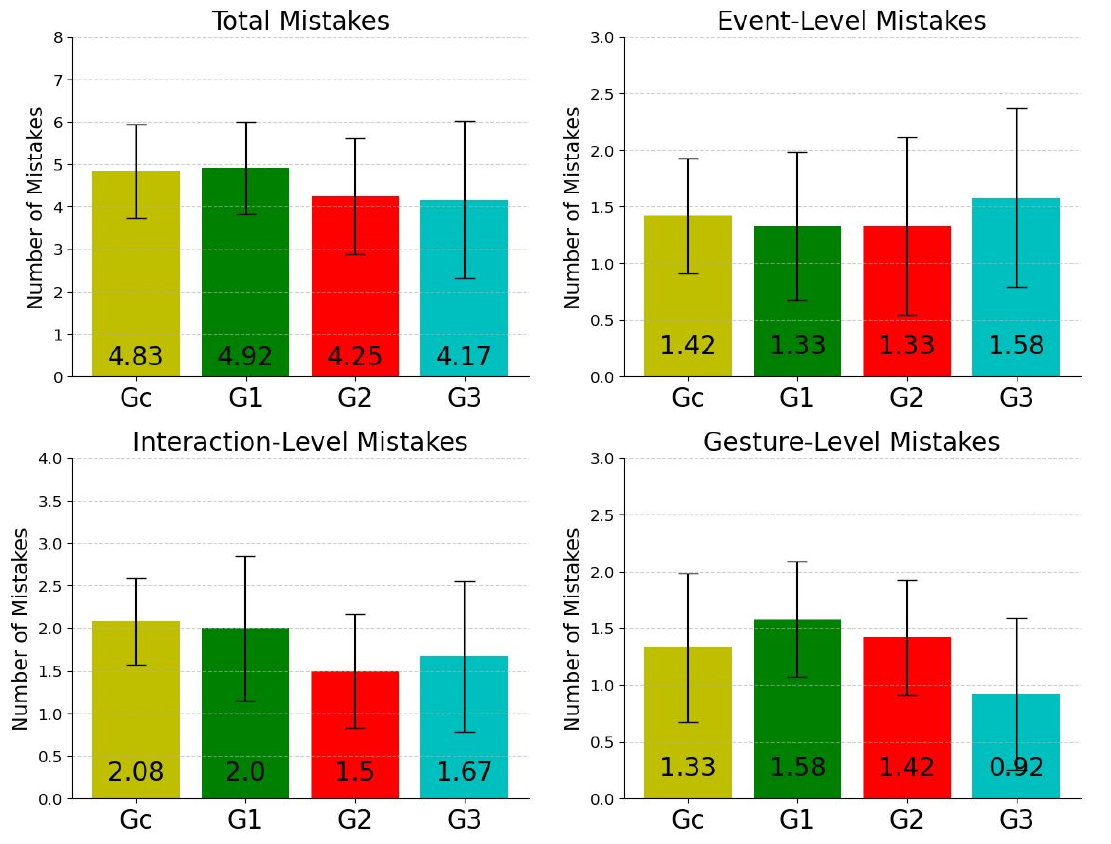}
    \caption{\textbf{Average Mistakes made per participant} during the testing phase in different levels. (error bars represent standard deviations)}
    \label{fig:mistakes}
\end{figure}

We evaluate the remaining four hypotheses \textbf{H2, H3, H5}, and \textbf{H6} in this section.
The participant's performance during the testing phases with respect to the average total task time  is shown in ~\autoref{fig:testing_time} (\textbf{H2} and \textbf{H5}).
The number of mistakes of three different levels made during testing phase is demonstrated in ~\autoref{fig:mistakes} (\textbf{H3} and \textbf{H6}).

The Friedman test shows significant differences among the groups in testing time ($\chi^2(3)=15.9, p=0.001$)
The average testing time of the participants from {G3}($M=4.31, SD=0.35$) is significantly lower than all three other groups, namely {Gc}($M=5.20, SD=0.69, p=0.002, Z=-2.824$), {G1} ($M=4.92, SD=0.47, p=0.001, Z=-2.981$), and {G2} ($M=4.61, SD=0.24, p=0.036, Z=-1.804$). This proves \textbf{H2 and H5}.
The average testing time of the participants from {G1} is  {not significantly different from} those from {G2} ($p=0.058, Z=-1.5689$).

We also evaluated the number of mistakes made by the participants during the testing phase.
The Friedman tests show no significant difference among the groups in total mistake numbers($\chi^2(3)=3.3, p=0.348$), event-level mistake numbers ($\chi^2(3)=0.5, p=0.919$), interaction-level mistake numbers ($\chi^2(3)=5.825, p=0.120$), and gesture-level mistake numbers ($\chi^2(3)=4.525, p=0.210$). This shows that both \textbf{H3 and H6} fails.

\subsection{Subjective Feedback}
\subsubsection{Likert-Scale Questionnare}
\begin{figure}[htp]
    \centering
    \includegraphics[width=\linewidth]{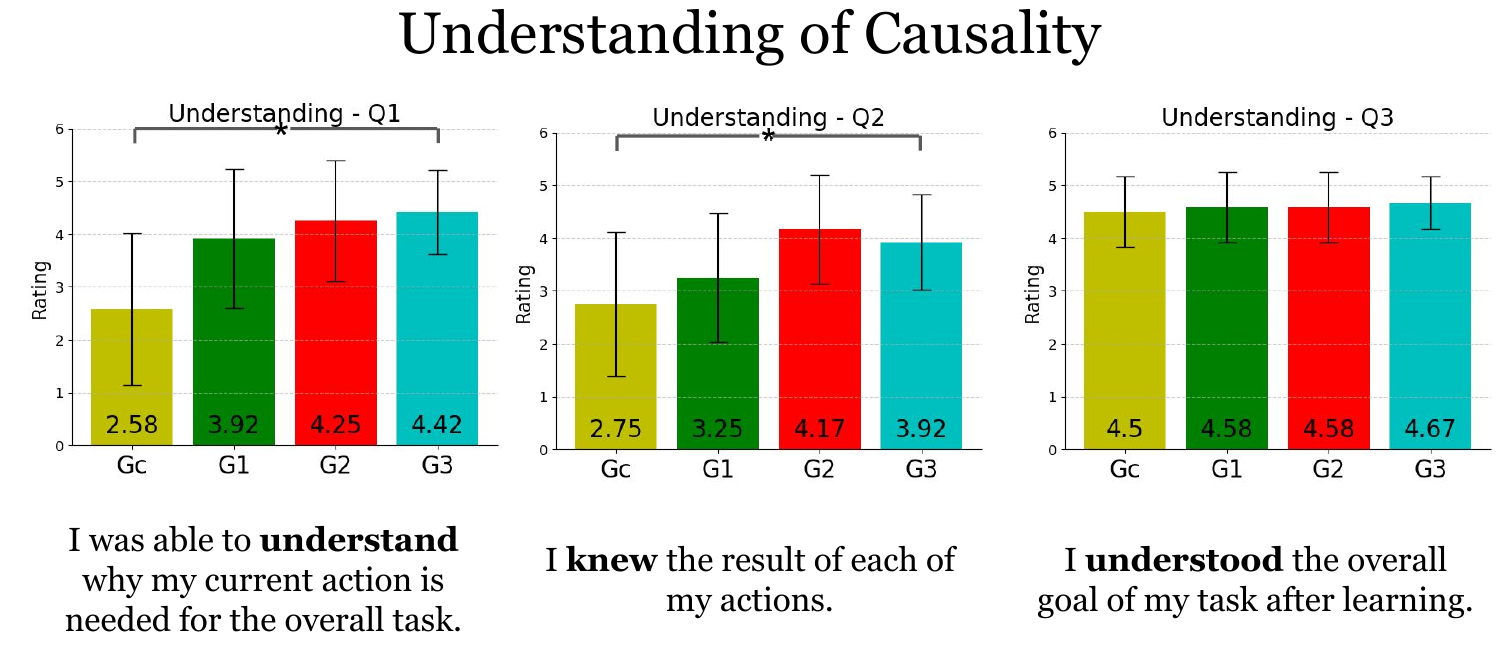}
    \caption{Subjective User Rating on their \textbf{Understanding of the task}, overall or in detail. The results are average scores by group from a 5-point Likert-scale questionnaire. $**=p<.005,*=p<.05$, error bars represent standard deviations.}
    \label{fig:Understanding_likert}
\end{figure}
 
\begin{figure}[htp]
    \centering
    \includegraphics[width=\linewidth]{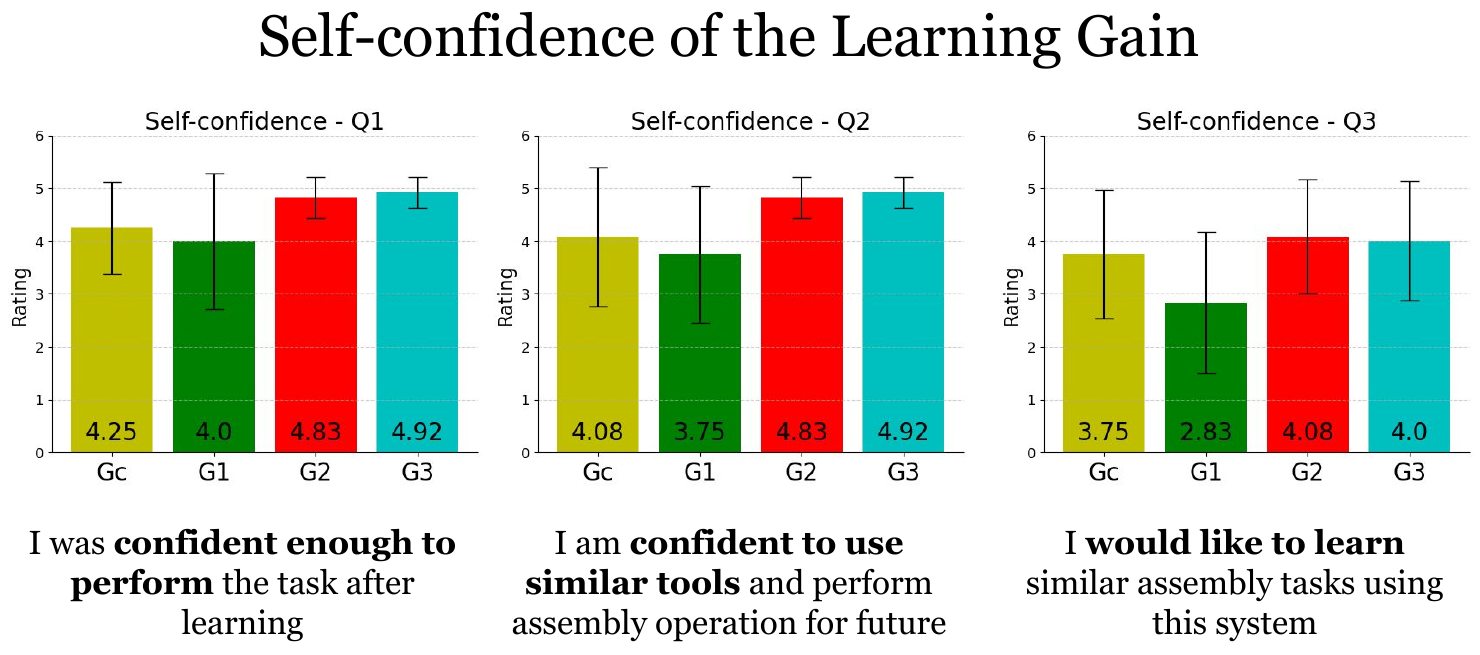}
    \caption{Subjective User Rating on their \textbf{Self-confidence in the learning gain}. The results are average scores by group from a 5-point Likert-scale questionnaire. No significant difference was found. Error bars represent standard deviations.}
    \label{fig:confidence_likert}
\end{figure}

\begin{figure}[htp]
    \centering
    \includegraphics[width=\linewidth]{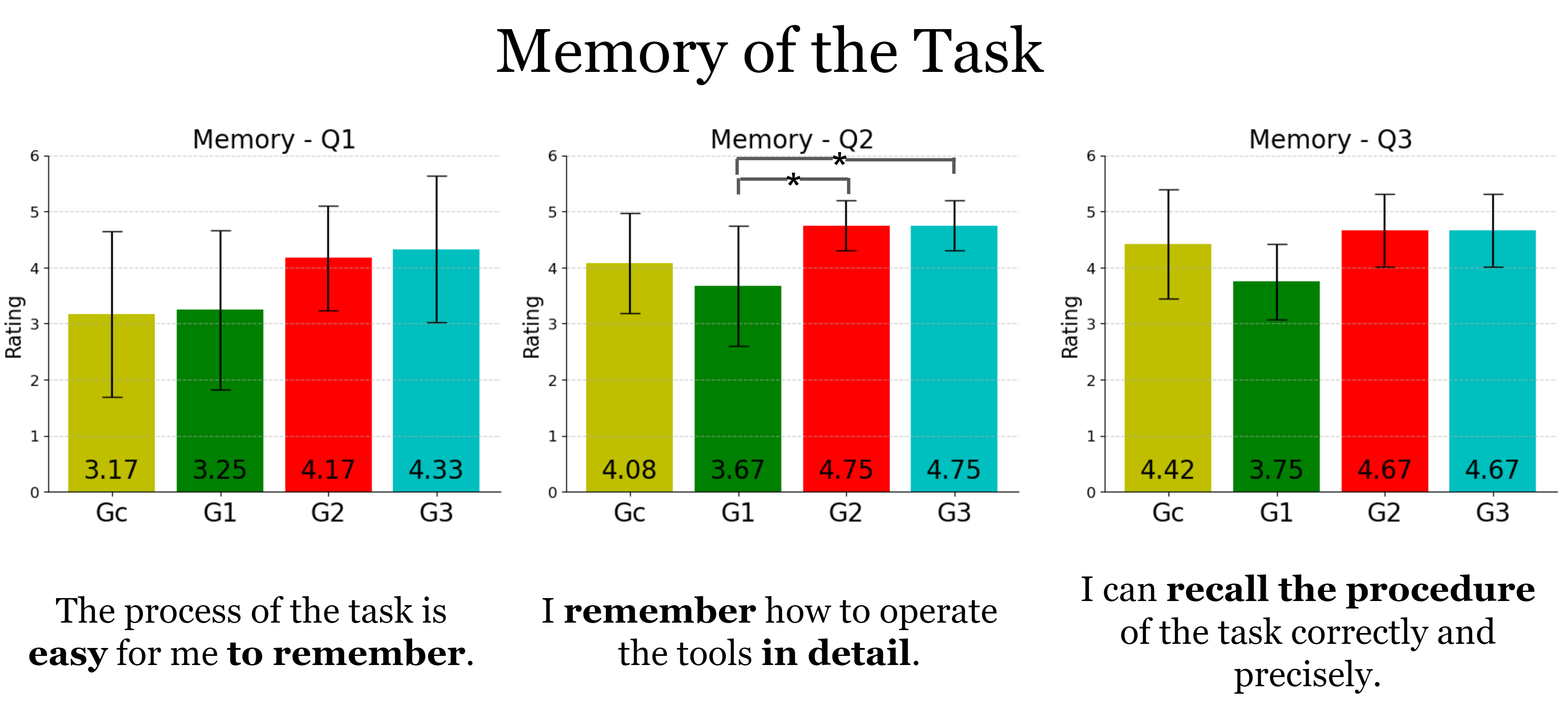}
    \caption{Subjective User Rating on their \textbf{Memory of the learning content}. The results are average scores by group from a 5-point Likert-scale questionnaire. $**=p<.005,*=p<.05$, error bars represent standard deviations.}
    \label{fig:Memory_likert}
\end{figure}

We qualitatively evaluate user feedback using a 5-point Likert Scale Questionnaire consisting of 3 major categories Understanding, Motivation, and Memory. We performed the Friedman test to compare all the questions in the respective categories. The results are presented below ~\autoref{fig:Understanding_likert}.

\begin{itemize}
    \item \textbf{Understanding.} The Friedman test shows a significant difference between all four groups for Q1($\chi^2(3)=8.325, p=0.040$) and Q2($\chi^2(3)=8.05, p=0.044$) but not Q3($\chi^2(3)=0.175, p=0.98$). We performed pairwise comparisons among the groups for Q1 and Q2.  {For Q1, {G3} ($M = 4.42, SD = 0.75$) understands the importance of the task better as compared to {Gc} ($M = 2.58, SD = 1.24, p = 0.005, Z = -2.548$).}  {G3} ($M = 3.92, SD = 0.9$) also knows better the outcome of their actions beforehand as compared to {Gc} ($M = 2.75, SD = 1.35, p = 0.038, Z = -1.765$) ~\autoref{fig:Understanding_likert}. 

    \item \textbf{Self Confidence.} The Friedman test does not reveal any significant difference for all the questions Q1($\chi^2(3)=5.725, p=0.125$), Q2 ($\chi^2(3)=6.775, p=0.079$), and Q3 ($\chi^2(3)=4.675, p=0.197$) in self-confidence ~\autoref{fig:confidence_likert}.
    
    \item \textbf{Memory.} The Friedman test shows significant variation in Q2($\chi^2(3)=7.041, p=0.029$) but not in Q3 ($\chi^2(3)= 5.975, p=0.112$) and Q1 ($\chi^2(3)=5.875, p=0.117$). {G3} (M = 4.75, SD = 0.45) users can remember to operate the tools better than {G1} ($M = 3.67, SD = 1.07, p = 0.0007, Z = -2.310$). Also, {G2} can better remember the tool use as compared to {G1} ($M = 3.67, SD = 1.07, p = 0.007, Z = -2.446$).     
\end{itemize}

\subsubsection{NASA TLX}

\begin{figure}[htp]
    \centering
    \includegraphics[width=\linewidth]{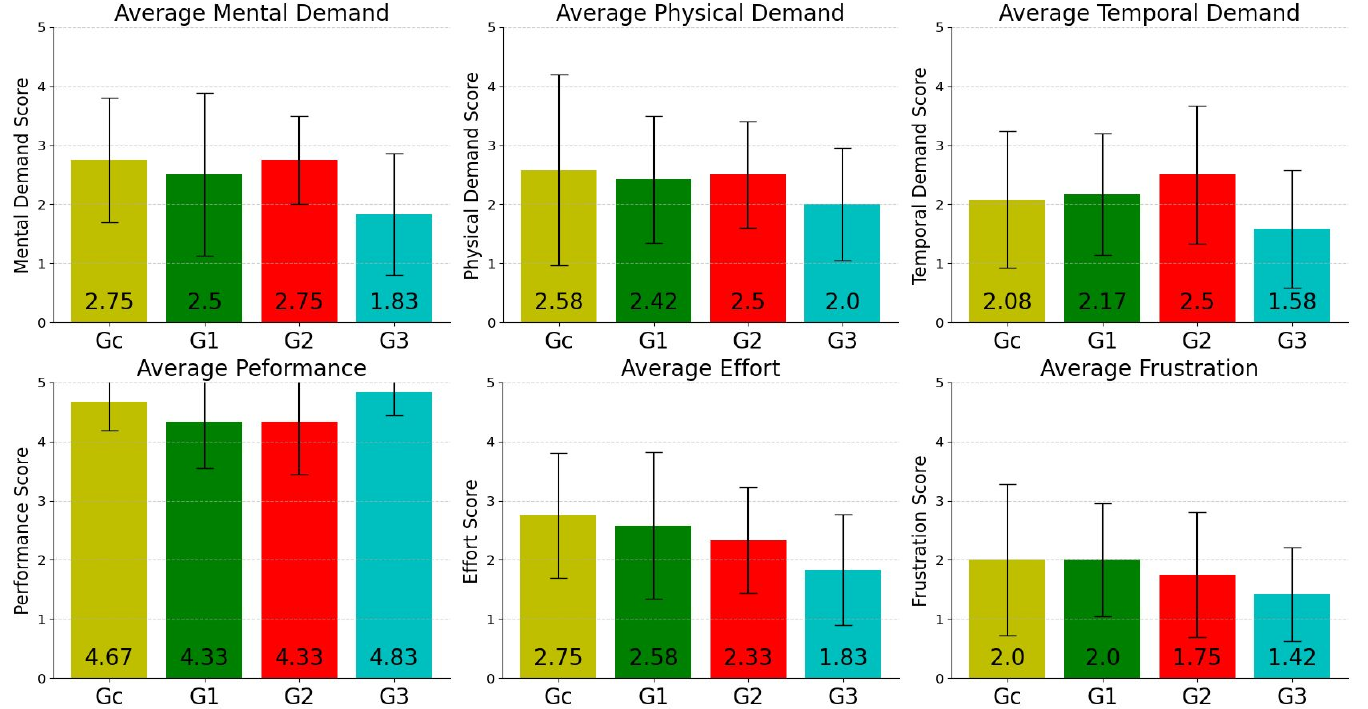}
    \caption{\textbf{NASA TLX results}. Error bars indicate the standard deviations. We found no significant difference among the groups.}
    \label{fig:nasatlx}
\end{figure}

We used NASA TLX's responses to evaluate the perceived subjective mental workload experience.
The findings are present in ~\autoref{fig:nasatlx}. After confirming the non-normality with a Shapiro-Wilk test, a Friedman test was conducted on the results obtained from an unweighted NASA TLX to compare all the groups.
None of the metrics showed significance: Mental Demand ($\chi^2(3)=3.875, p=0.275$), Physical Demand ($\chi^2(3)=1.8, p=0.614$), Temporal Demand($\chi^2(3)=4.225, p=0.238$), performance ($\chi^2(3)=2.625, p=0.453$), Effort($\chi^2(3)=3.525, p=0.317$) and frustrations($\chi^2(3)=2.125, p=0.546$).
We further do not perform any adjustments in p-values for multiple comparisons, as there is no significant result as shown in ~\autoref{fig:nasatlx}.

\subsection{Research Question Analysis}

\begin{itemize}
    \item \textbf{RQ1}: We observed that \textbf{H1} and \textbf{H3} did not yield successful outcomes, whereas \textbf{H2} was confirmed to be accurate.
     {\textbf{H1} posited that the learning time in {G3} would be less as compared to {Gc}.
    Yet, the learning time in {G3} is significantly higher than that of {Gc}.}
    After failing of two hypotheses we further analyzed more results.
    To explore performance improvements during the learning phase, we incorporated an additional metric based on the number of tutorial revisits~\cite{cao2020exploratory}.
    Interestingly, the average revisits for {Gc} ($M=3.17, SD=0.72$) were significantly higher compared to {G3} ($M=2.33, SD=0.49, p=0.018, Z=-2.090$), as depicted in ~\autoref{fig:learning_results}.
    This shows the difficulty in learning the content.
    Regarding \textbf{H3}, it was inconclusive due to the absence of a significant difference.
    We conclude that the absence of significance in the number of mistakes is due to the similar mistakes made during testing, which led to close counts of mistakes across the groups at all three levels.
     {The rejection of H1 and H3 along with the confirmation of H2 suggest the answer to RQ1 is that visualizing causality in task learning results in a shorter performance time in testing with the compromise of more time spent in learning while there is no evidence of fewer errors while performing the task.
}

    Moreover, the NASA TLX workload assessment results did not yield significance across all four groups.
    Through a combination of our original hypotheses and additional analyses, we can deduce that the inclusion of causal information improves the quality of the learning gain and thus enhances the performance of the participants in the same task.
    However, additional causal information induces more content to learn, resulting in a longer learning time.
    Yet, it is worth noticing that despite the extra content to learn, participants from {G3} revisited the tutorials less than those from {Gc}, implying a better immediate learning gain with an understanding of the causal information in the tasks.
    
    \item \textbf{RQ2}: We can see from the results that \textbf{H4} is rejected although {G1} did take the most time which corresponds to event-level causality.
    \textbf{H5} was proven valid whereas \textbf{H6} is not supported.
    From further analyzing the data we found that the reason behind the failure of \textbf{H6} is the same as discussed for \textbf{H3} in the above paragraph for the mistakes made in the task.
     {To answer RQ2, we conclude from our study that among all three levels, visualizing only event-level causality introduces the most time spent in learning while visualizing causality at the interaction level costs the least time in learning.
    In addition, learning with gesture-level causality visualization results in the least completion time post-learning, while there is no evidence that learners would make fewer mistakes post-learning.}
\end{itemize}
\section{Discussion}
In this section, we summarize key results and provide insight drawn from our analysis and post-study interviews.

\subsection{Better Learning Results with Intention-driven Causality}
The average learning time of {Gc} is significantly less than that of {G3} indicating that participants are spending more time learning additional information about the intention-driven causality of the assembly task.
However, the significantly high number of revisits from {Gc} indicates that with the knowledge of causality, the learners understand the tasks with fewer times referring to the instructions because they get insights from the additional causality information.
In terms of the testing performance, participants from {G3} perform significantly faster than participants from {Gc}, with fewer total mistakes made during their demonstrations.
The ratings from the post-study questionnaire \textbf{Understanding - Q1} and \textbf{Understanding - Q2} show that participants from {G3} give significantly higher ratings than those from {Gc} regarding their understanding of the consequences of and intention of their actions in question, which leads to better performance of the task \textit{"I knew I needed to clamp the base tight enough when I saw my next step was to hit it with a hammer ({G2}, P4)"}.
This finding aligns with the conclusion from the prior work~\cite{gardiner2011guided, luchkina2018more, buehner2009causal, liu2022precueing} that in observational causal learning, the understanding of the causality is key to the learning performance and learning outcome.



\subsection{Levels of Details: the Deeper, the Better} \label{Levels of Details}

 {From our cross-comparisons among {G1}, {G2}, and {G3}, we notice that both {G2} and {G3}  have significantly less learning time than {G1}.}
The higher learning time of {G1} comes from the fact that showing only event-level instruction makes it hard for the learners to memorize the procedure and results in the learners more frequently referring to the instructions.
\textit{"The instruction is too long for me to remember, and I have to rewind to the part I am doing ({G1}, P3)."}
 {The feedback from \textbf{Memory - Q2} indicates that without learning the interaction and gesture/affordance levels of intention-driven causality, {G1} has a poorer understanding of how to operate the tool.
A similar conclusion can be drawn from \textbf{Memory - Q3} where {G1} gives lower ratings on their performance than {G2} and {G3} regarding their precision.}


Prior works~\cite{Lindlbauer2017Context, huang2021adaptutar, Tatzgern2016Adaptive} have shown that the more details provided to the users of the MR system, the higher cognitive load is forced on the users, resulting in decreasing learning performance.
Yet, with details added in our study, we do not observe a significant increase in learning time or a significant difference in users' subjective confidence about the task they have learned (\textbf{Self-confidence - Q1,2,3}).
 {To sum up, we conclude that visualizing gesture-level causality information in manual task learning in MR provides the learners with the best understanding and memorization of the tasks and the highest post-learning time performance in the tasks.
However, gesture-level causality visualization in manual task learning is not the best practice when efficiency in learning is the priority, because it results in more learning time compared to no-causality-visualization.}

\subsection{In-place 3D visualization}
 {In our study, we employed a hybrid approach, combining 2D videos and 3D animations, to present content to learners about the tasks}~\cite{chidambaram2021processar}.
 {We utilized a hierarchical representation in our visualizations to align with the hierarchical mental models that humans naturally employ to decompose and understand tasks.
Participants are prompted to look at 3D visualization for the current task, and then look at the hierarchical 2D visualization for complete causal information.
The different placements of two visualizations can cause discontinuity in participants' learning experiences.

A potentially more promising way to visualize both pieces of information and avoid discontinuity is via full 3D animations fixed at the objects showing current tasks followed by the causal tasks}~\cite{henderson2010exploring}. 
 {Additionally, other forms of visual cues}~\cite{liu2022precueing, liu2022adaptive, Liu2022Rotation} { such as arrows, bounding boxes around the objects, and textual description can be integrated to provide a more designated learning purpose.
Learners can better anticipate the implications of their actions within the 3D space}~\cite{cao2020exploratory} {, leading to deeper learning}

\section{Future Work and Limitation}
\subsection{Future Work}
Given the insights from the results, we would like to provide recommendations for further integration of causality and intention in future MR manual task learning systems.

\subsubsection{Visualization of Causality}
One of the less satisfying results from our study is the compromise in learning time learners have to make when learning additional information about causality and intention of the task, as mentioned in ~\autoref{Levels of Details}.
There is always a trade-off between the content's volume and the time spent learning.
For example, "\textit{I could understand that my action would reflect the task, but it took me a while to realize how (\textbf{G4}, P11.)}"
Therefore, we suggest that natural, user-friendly, and dedicated visualization techniques can be applied to causality and intention in MR.
Take the scenario of making an omelet for example.
Should the learners set the heat too high when preheating the pan, a 3D animation of an omelet being overcooked can be overlaid on the pan in MR.
Such dedicated visualizations can intuitively acknowledge the user's causality within the tasks without significantly increasing the time and cognitive cost.
Yet, such visualizations' design, formulation, and generalization remain challenging.
One promising direction is to utilize human attention during an action.
State-of-the-art artificial intelligence algorithms can predict human attention based on their actions ~\cite{yang2020attention, baee2021medirl, ZELINSKY2020231}.
By predicting the engagement based on learners' interaction with the objects and better positioning the visualization, it is possible that the learner can naturally perceive knowledge without a high cognitive cost in transitioning. 

\subsubsection{Dynamic Causality} \label{Dynamic Causality}
Showing merely one future causally connected to the present results in neglected details is not optimal for learning.
Therefore, we propose enhancing the representation of causality and human intention in manual task learning by \textit{\textbf{dynamically modeling causality}}.
In other words, the learner sees how the causally connected futures are affected by their actions in the present.
Furthermore, by showing the future dynamically, the learners can visualize and understand a broader range of causality, providing a complete understanding of the task.

Another motivation for \textit{\textbf{dynamically modeling causality}} is the suggestion from some psychology research ~\cite{chillarege2003learning, Andrew2004How} that humans can learn not only from correct instruction but also from the adverse effects of their actions.
By showing the possible future failure caused by their current mistake, learners may re-evaluate their reasoning and gain a deeper understanding of the task's principles.
We anticipate promising results from incorporating a dynamic representation of the future in MR manual task learning systems.

\subsubsection{Comparing methods for task decomposition}

 {In our study, the tasks are decomposed based on three levels of causal hierarchy derived from cognitive psychology}~\cite{RIVA201124}.
 {We employed an object-level segmentation of tasks, where causality is determined based on the relationships between tasks that share common objects} ~\cite{fender2022causality}.
 {While we focus on identifying and categorizing the levels of causality in manual tasks and design experiments specifically for causality, we acknowledge that other forms of task decomposition methods can be used to segment the task and influence the learning results.
Vitally, prior work}~\cite{heiser2004identification} { has shown that the information shown in each visualization has a great impact on the performance of the users following the visualization.
We envision future works to incorporate our study findings into state-of-the-art task decomposition methods to optimize task learning with interactive systems}~\cite{bowen2021task}.

\subsubsection{Causality addressing Error Cascades in Task Execution}

{Initial errors can lead to a chain of mistakes which will impact subsequent task performance and reduce the overall performance. 
Visualizing causal links within the task can help users understand how a current error influences future steps and the final outcome of the task.
This type of error visualization highlights \textit{why} each step is important and how it contributes to the overarching goal of task completion. 
Future directions include implementing real-time error detection methods, employing adaptive visualizations to illustrate the cascading effects of errors on subsequent actions, and providing corrective feedback.
These can improve task learning by mitigating the negative impact of errors, enhance understanding, and improve learning gains.}

\subsection{Limitation}

\subsubsection{Pre-test for knowledge gathering}
 {Our study considered only post-test knowledge testing with a two-fold reasoning.
First, participants were new to the task and unfamiliar with our MR system of causality, and hence we assumed a uniform baseline of knowledge.
Second, considering the manual task procedure and operations involved in our study are monotonous, we forwent a pre-test to avoid potential learning effects that could mitigate the findings in our post-test.
Therefore, we acknowledge that an ideal pre-test could have been conducted to assess the participants' prior knowledge of the manual tasks in our study, and therefore could have constituted a more comprehensive analysis of the learning gain.}

\subsubsection{Test-bed Limitations}
Our workflow for generating 3D instructions and segmenting tasks relies on the tracking 6 DoF of the objects and 3D gestures of the hands.
Hence, we suffer from the same limitations state-of-the-art tracking algorithms face, such as occlusion, blurry images/feed, small-size tracking, and computation efficiency.
Therefore, relying entirely on camera-based systems and generalizing the workflow remains challenging to obtain high-quality 3D content. However, compromises can be made, such as using 6 DoF sensors, multiple cameras, or markers-based computer vision to enhance the accuracy.
One may also be concerned about tracking non-rigid objects, such as flexible wires and cables.
However, with advances in computer graphics, state-of-the-art vision-based algorithms can also obtain a 3D model of non-rigid objects~\cite{lakhili2022rigid}, which is encouraging.
Finally, we would like to state that the methodology applied here for tracking does not relegate the comprehension of causality and intention matter in manual task learning.

To test our hypothesis, we have implemented a systematic instance of the workflow that is capable of preserving the causal relation and intention and delivering them to the learners.
Along with this implementation, we designed a use case for an assembly task that involves interaction with multiple objects and causally related events.
Even though the use case is designed based on real-world hand-object interactions and assembly tasks, it is yet a test bed.
We want to acknowledge that this study aims to evaluate the implementation of our theory, and the results are analyzed to justify our approach and to bring insights into the potential contribution the understanding of causality and intention can make to manual task learning in the MR community.
We also identified a blemish in our experiment setup, as the task in the user study might have been simplified, the mistake numbers across the groups did not show significant differences, which failed to support \textbf{H3} and \textbf{H6}.
We anticipate a more rigorous setup and metrics to evaluate the learning gain of understanding causality in terms of mistakes made during the tasks.

Finally, we propose our work as an elicitation for future MR manual task learning systems.

\subsubsection{User Study Sample Size}

{Following the convention of 12 participants per group for usability and quantitative evaluation in HCI}~\cite{caine2016local} , {we included 12 participants per condition in our study. 
We analyzed the effect sizes post hoc. 
We found that our sample size of 12 produced varying levels of effect sizes from small effect to large effect. 
We specifically found large effect sizes for metrics such as learning time, number of repeats, and completion time in the testing phase. 
We found small effects for the mistake calculation.
For NASA TLX and subjective feedback, the effect again varies. 
As we have results for mixed effect sizes, sample size may have been small to detect smaller or medium effect sizes reliably. 
While the current sample size was sufficient for detecting the observed differences in large effect size, increasing the number of participants could further improve the precision of effect size estimates and allow for the detection of small or medium effect.
This adjustment will provide greater statistical power, reduce variability, and confirm the reliability of observed effects. 
}
 
\section{Conclusion}
In this paper, we studied the effect of understanding causality and intention in manual task learning in MR.
We studied the importance of causal understanding in the learning process and present the concept of \textit{\textbf{Intentional-driven Causality for manual task learning in MR}}, consisting of three layers of causality information in a manual task, namely, event level, interaction level, and gesture level.
We then built a test bed to study the effect of different layers of causality on the learning gain of an assembly task.
We conducted a thorough user study with 48 users, aiming to answer the research questions of whether and how visualizing causality helps manual task learning in MR, as well as the effect of different levels of causality in manual task learning.
With the quantitative and qualitative results from the study, we discuss and attribute the improvement in learning shown in the analysis, as well as all other phenomena and insights, worthy of more research.
We conclude that participants perform the assembly task better after learning the causality information and gesture-level causal information grants the best performance among all three levels.
Besides, we also draw other conclusions on more findings from the study.
Finally, we envision the future work from our study and the concept of \textit{\textbf{Intentional-driven Causality for manual task learning in MR}} to help the community create natural, efficient, and learner-friendly manual task learning methodologies with MR.

\section*{Acknowledgement}
This work is partially supported by the NSF under the Future of Work at the Human-Technology Frontier (FW-HTF) 1839971 and NSF Partnerships for Innovation Technology Transfer (PFI-TT) 2329804. We also acknowledge the Feddersen Distinguished Professorship Funds and a gift from Thomas J. Malott. Any opinions, findings, and conclusions expressed in this material are those of the authors and do not necessarily reflect the views of the funding agency.
 
%
\bibliographystyle{IEEEtran}
\bibliography{texes/reference}

\begin{IEEEbiography}[{\includegraphics[width=1in,height=1.25in,clip,keepaspectratio]{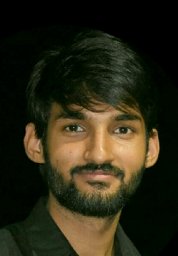}}]{Rahul Jain}
Rahul Jain is a Ph.D. student in the School of Electrical and Computer Engineering at Purdue University. He received his Master’s in Electrical and Computer Engineering at Purdue University and a Bachelor’s in Civil Engineering at the Indian Institute of Technology (IIT), Patna. His current research focuses on the area of Computer Vision, Machine Learning, and human-computer interactions utilizing AR/VR.
\end{IEEEbiography}

\begin{IEEEbiography}[{\includegraphics[width=1in,height=1.25in,clip,keepaspectratio]{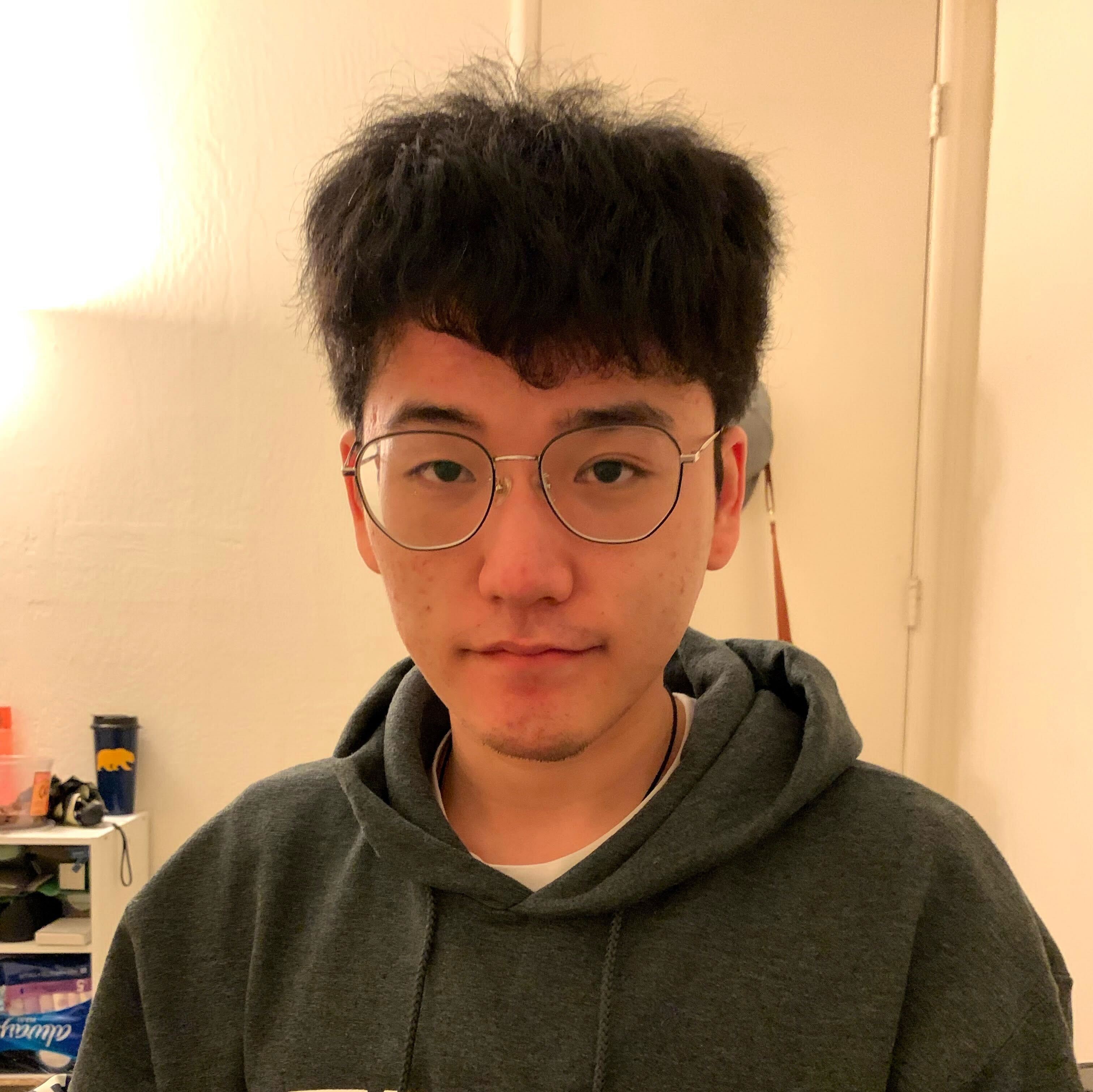}}]{Jingyu Shi}
Jingyu Shi is a Ph.D. candidate in the Electrical and Computer Engineering Department of Purdue University.
He obtained his Master of Science in Electrical and Computer Engineering in 2020 at Georgia Institute of Technology and his Bachelor of Engineering in Instrument Science at Beihang University.\end{IEEEbiography}

\begin{IEEEbiography}[{\includegraphics[width=1in,height=1.25in,clip,keepaspectratio]{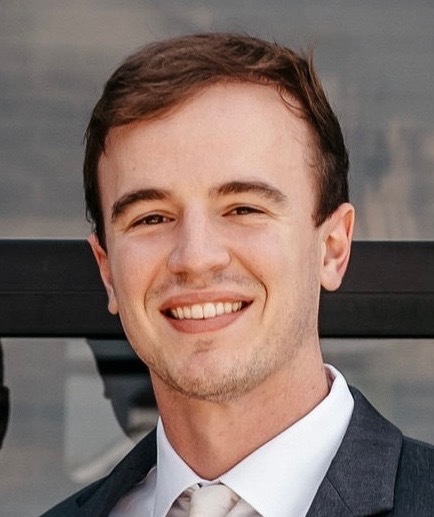}}]{Andrew Benton}Andrew Benton is an Applications Engineer at Mecademic, where he designs solutions for industrial micro-automation. He received his Master of Science in ECE from Purdue University and his Bachelor's degree in Computer Engineering from the University of Denver.
\end{IEEEbiography}

\begin{IEEEbiography}[{\includegraphics[width=1in,height=1.25in,clip,keepaspectratio]{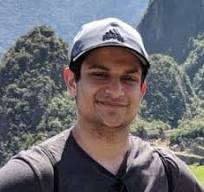}}]{Moiz Rasheed}
Moiz Rasheed is a Calibration Software Engineer at IonQ. He received his Masters of Science and his Bachelors in ECE at Purdue University. He is now working on software to make quantum computers run more reliably and increase uptime.
\end{IEEEbiography}

\begin{IEEEbiography}[{\includegraphics[width=1in,height=1.25in,clip,keepaspectratio]{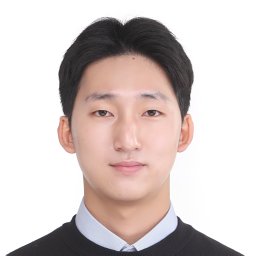}}]{Hyungjun Doh} Hyungjun Doh is a Master's student in the Electrical and Computer Engineering Department at Purdue University. He earned his Bachelor's degree in Computer Engineering from Purdue University. His research focuses on Computer Vision, with a particular emphasis on 3D reconstruction for Human-Object Interaction.
\end{IEEEbiography}

\begin{IEEEbiography}[{\includegraphics[width=1in,height=1.25in,clip,keepaspectratio]{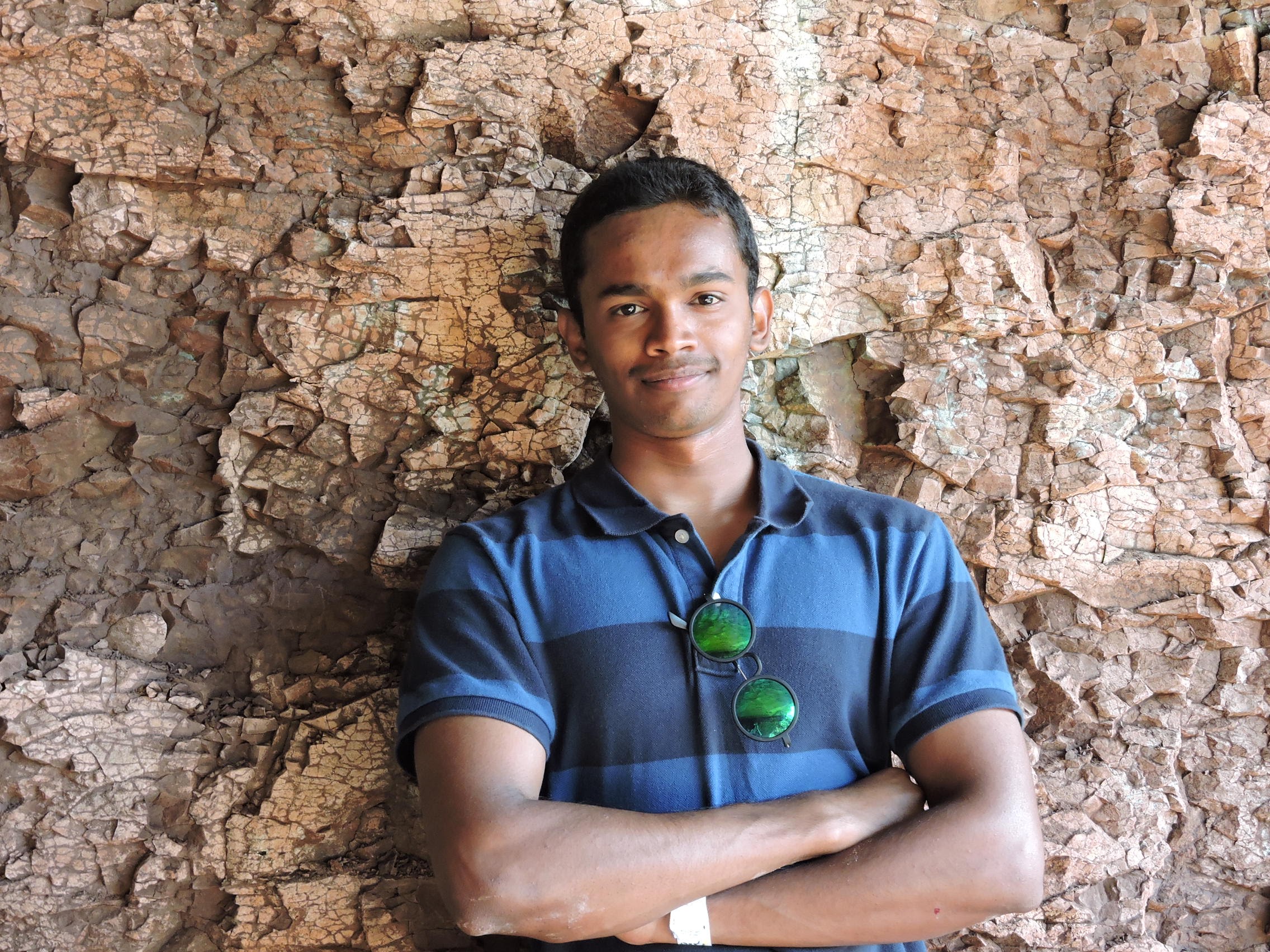}}]{Subramanian Chidambaram}
Subramanian Chidambaram is an applied scientist at Amazon Web Services specializing in intersectional research on human-computer interaction (HCI) and artificial intelligence (AI). He holds a Ph.D. in Mechanical Engineering from Purdue University, where he was a researcher at the C Design Lab. His doctoral work focused on designing, developing, and evaluating novel 3D interfaces and interaction techniques to advance Extended Reality (XR) applications. His contributions to this publication were made during his time as a Ph.D. student at Purdue.
\end{IEEEbiography}

\begin{IEEEbiography}[{\includegraphics[width=1in,height=1.25in,clip,keepaspectratio]{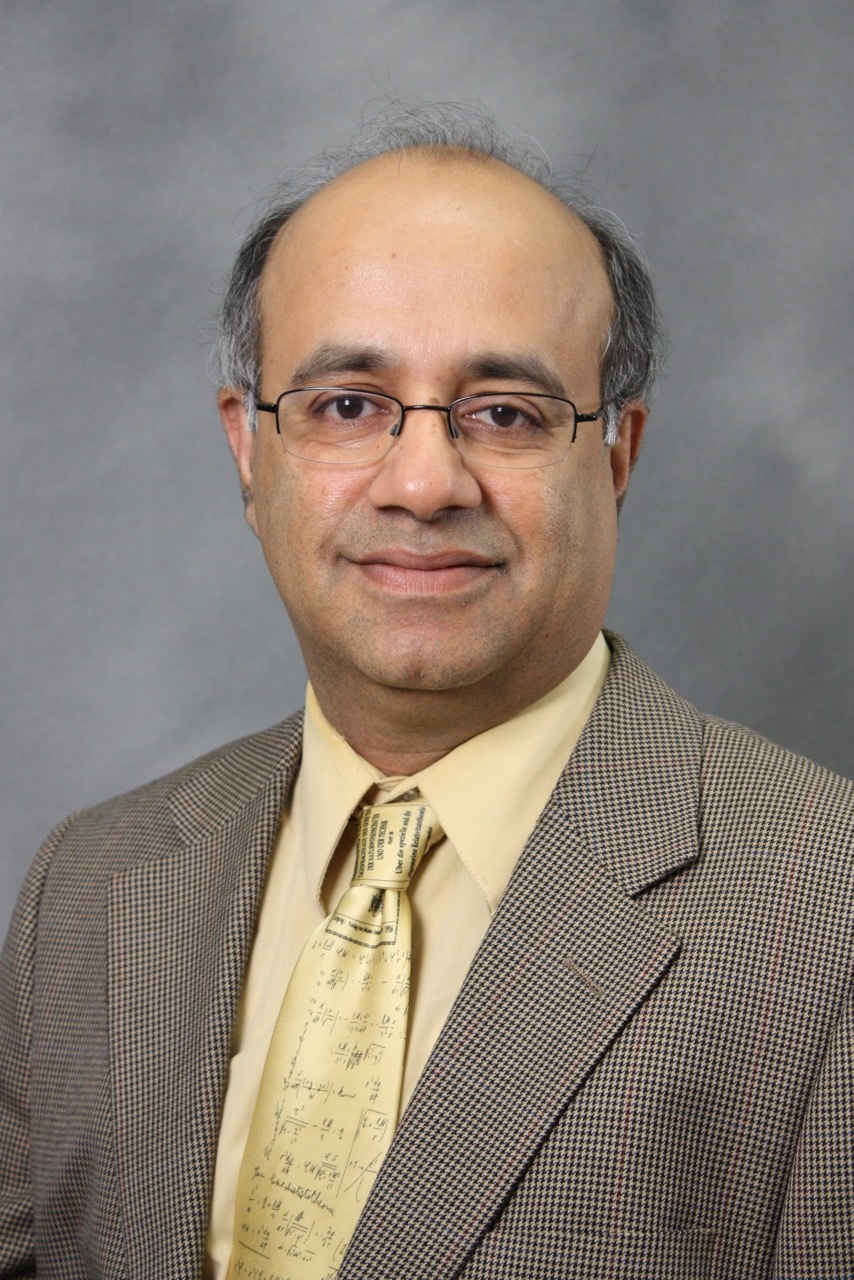}}]{Karthik Ramani}
Karthik Ramani is the Donald W. Feddersen Distinguished Professor of Mechanical Engineering at Purdue University, with appointments in Electrical and Computer Engineering and Education. His research focuses on augmenting human capabilities through virtual and physically blended tools and developing AI-driven XR authoring systems. He has co-founded VizSeek, the first commercial visual search company, and ZeroUI for educational robotics. A former Visiting Professor at Oxford and Stanford, Ramani has received Purdue’s Research Excellence and Commercialization Awards and also coaches Purdue’s nationally ranked table tennis team.
\end{IEEEbiography}
\end{document}